\documentclass[12pt]{iopart}
\usepackage{cite}
\usepackage{color}

\usepackage{amssymb}
\usepackage{graphicx}

\usepackage{graphicx}
\usepackage[hang,small,bf]{caption}
\usepackage{amsfonts}

\usepackage[subrefformat=parens]{subcaption}

\usepackage{cases}
\begin{document}

\title{Velocity control for improving flow through a bottleneck}

\author{Hiroki Yamamoto$^1$, Daichi Yanagisawa$^1$ and Katsuhiro Nishinari$^2$}

\address{$^1$ Department of Aeronautics and Astronautics, School of Engineering, The University of Tokyo,\\ 7-3-1 Hongo, Bunkyo-ku, Tokyo 113-8656, Japan}
\address{$^2$ Research Center for Advanced Science and Technology, The University of Tokyo,\\ 4-6-1 Komaba, Meguro-ku, Tokyo 153-8904, Japan}

\ead{hirokiyamamoto@mbr.nifty.com}

\vspace{10pt}
\begin{indented}
\item[]January 2016
\end{indented}

\begin{abstract}
A bottleneck can largely deteriorate the flow, such as a traffic light or an on-ramp at a road. To alleviate bottleneck situations, one of the important strategies is to control the input rate to suit the state of the road.  In this study, we propose an effective velocity control of traveling particles, in which the particle velocity depends on the state of a bottleneck. To analyze our method, we modify the totally asymmetric simple exclusion process (TASEP) and introduce a slow-to-start rule, which we refer to as controlled TASEP in the present paper. Flow improvement is verified in numerical simulations and theoretical analyses by using controlled TASEP.
\end{abstract}

%
%
%
%
%

\section{Introduction}
Recently, researchers in non-equilibrium physics have become interested in systems of active matter. These systems, with their remarkable self-propulsion features, have been vigorously investigated in numerical simulations, experiments, and theories~\cite{RevModPhys.85.1143, PhysRevE.90.012701, PhysRevE.91.032117}. Despite their simplicity, such systems describe a wide range of phenomena far from thermal equilibrium. Among the various models of active matter, the so-called asymmetric simple exclusion process (ASEP) has attracted much attention as a minimal model.

The ASEP was pioneered by MacDonald and Gibbs~\cite{macdonald1968kinetics, macdonald1969concerning}, who sought to describe the kinetics of protein synthesis. Since exactly solving the model with open boundary conditions~\cite{1751-8121-40-46-R01, 0305-4470-26-7-011, PhysRevE.59.4899}, researchers have applied the ASEP and its extensions to diverse systems such as vehicular traffic~\cite{RevModPhys.73.1067}, molecular motor traffic~\cite{parmeggiani2003phase, nishinari2005intracellular,chou2011non, pinkoviezky2013transport, pinkoviezky2014traffic,appert2015intracellular}, exclusive queuing processes~\cite{arita2010exclusive, yanagisawa2010excluded, arita2014dynamics}, and market systems~\cite{willmann2002exact}. 

A simplified version of ASEP called the totally asymmetric simple exclusion process (TASEP), where particles are allowed to hop unidirectionally (left to right in the present paper), is popularly applied in vehicular traffic modeling for its conceptual simplicity. 
Traffic management strategies are important for mitigating traffic jams and improving traffic flows at bottlenecks. Therefore, they have been extensively studied by TASEP in recent years~\cite{PhysRevE.91.062818,1742-5468-2014-10-P10019, 1742-5468-2009-02-P02012, 1742-5468-2012-05-P05008, 1742-5468-2008-06-P06009, PhysRevE.87.062818, Nagatani2016131}. 

In practice, traffic-management strategies are typically implemented by traffic lights. However, a traffic light can also behave as a bottleneck. Under the law of inertia, stopping in front of a traffic light can deteriorate traffic flow. To address this problem, many countries have installed roundabouts~\cite{wang2002modeling, PhysRevE.73.036101,PhysRevE.70.046132} to manage traffic flow at intersections. Roundabouts allow continuous vehicle movement and avoid deterioration in flow due to stopping. However, traffic lights remain effective for controlling traffic flows in cities, as roundabouts require a large area and can be detrimental in heavy traffic flows. Therefore, traffic flow management near traffic lights (and other bottlenecks) remains a worthwhile enterprise. Traffic lights have been actively investigated in many studies~\cite{popkov2008asymmetric,PhysRevE.89.042813,lammer2008self,salter1996highway}.

Many traffic-management strategies control an input ($\alpha$) or output ($\beta$) rate of particles (vehicles) according to the lane state, such as the lane density~\cite{1742-5468-2014-10-P10019, 1742-5468-2009-02-P02012, 1742-5468-2012-05-P05008, 1742-5468-2008-06-P06009,PhysRevE.87.062818}. In the TASEP approach of Woelki, the flow is improved by changing the input rate according to the lattice density~\cite{PhysRevE.87.062818}. In vehicular traffic, this method can be interpreted as a form of ramp metering~\cite{Gomes2006244, 881058}. However, these studies assume a constant hopping probability (rate) of particles. 

Our investigation adopts a different approach. We manage bottleneck flows by changing a particle velocity according to the state of a bottleneck. In the TASEP, if the right-neighboring site of a particle is empty, the particle moves to that site with a constant hopping probability $v$. We alter this rule depending on the state of the right boundary, i.e., the exit of the lattice. Specifically, $v$ is decreased only when the exit behaves as a bottleneck. This method can be interpreted as variable speed limit (VSL) control~\cite{papageorgiou2008effects} or jam-absorption driving ~\cite{nishi2013theory, Taniguchi2015304} in actual vehicular traffic. We note that related models, in which a hopping probability (rate) can differ among particles~\cite{evans1996bose, evans1997exact, krug1996phase} or sites~\cite{janowsky1992finite, janowsky1994exact}, have already been investigated.

Additionally, we introduce a slow-to-start (SlS) rule~\cite{doi:10.1142/S0218348X93000885,0305-4470-29-12-018, PhysRevLett.86.2498,sakai2006new, barlovic1998metastable, nishinari2004stochastic}, which is absent in the related works~\cite{1742-5468-2014-10-P10019, 1742-5468-2009-02-P02012, 1742-5468-2012-05-P05008, 1742-5468-2008-06-P06009,PhysRevE.87.062818}. The SlS rule represents the delay of restarting a blocked particle, accounting for the law of inertia and the reaction delay in actual situations. Note that the TASEP with the SlS rule can reproduce a metastable state, which is observed in real expressways~\cite{barlovic1998metastable}. 
Furthermore, by introducing additional rules, we can see a wide scattering area near the phase transition region from free flow to jam flow phase in the fundamental diagram~\cite{nishinari2004stochastic}.

A similar approach to our control method was developed by Nishi \textit{et al}~\cite{1742-5468-2011-05-P05027}. In their model, two open lanes merge into one lane. Particles, obeying the SlS rule, cannot change their lane and their hopping probability is decreased by particles in the other lane. The decreased hopping probability generates a spontaneous zipper merging, improving the particle flux at the merging point.

The remainder of the present paper is organized as follows. Sec. I\hspace{-.1em}I defines our modified TASEP, which we call controlled TASEP, and describes the modeled particle dynamics. In Sec. I\hspace{-.1em}I\hspace{-.1em}I and I\hspace{-.1em}V, we present and discuss the results of numerical simulations using controlled TASEP. The simulations are conducted to verify flow improvement by our method. The paper concludes with Sec. V.

\section{Model description}
\begin{figure}[htbp]
\begin{center}
\includegraphics[width=9cm,clip]{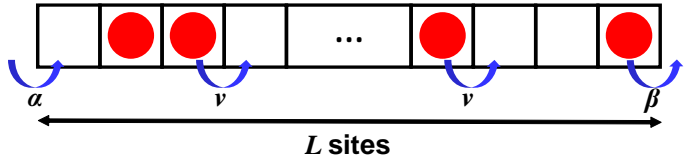}
\caption{Schematic of the original TASEP with open-boundary conditions.}
\label{fig:TASEP}
\end{center}
\end{figure}

The original TASEP with open-boundary conditions is defined as a one-dimensional lattice of $L$ sites, which are labeled from left to right as $i=0,1,......,L-1$ (see Fig. \ref{fig:TASEP}). Each site can be empty or occupied by a single particle. When the $i$th site at time $t$ is occupied by a particle, its state is represented as $n_i(t)=1$; otherwise, its state is $n_i(t)=0$. In the present paper, we adopt discrete time and parallel updating, which is generally used in traffic contexts. In parallel updating, the states of particles on the lattice are simultaneously determined in the next time step. Particles enter the lattice from the left boundary with probability $\alpha$, and leave the lattice from the right boundary with probability $\beta$. In the bulk, if the right-neighboring site is empty, particles hop to that site with probability $v$; otherwise they remain at their present site. If p
articles can hop to multiples sites with acceleration and deceleration under parallel updating in one time step, the TASEP becomes the Nagel-Schreckenberg (NS) model~\cite{nagel1992cellular}.

\begin{figure}[htbp]
\begin{center}
\includegraphics[width=9cm,clip]{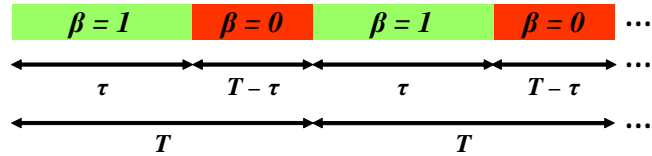}
\caption{Schematic of $\beta(t)$.}
\label{fig:exitmoshiki}
\end{center}
\end{figure}

Our controlled TASEP differs from the original TASEP in several ways. 
First, in our model, $\beta$ is not constant, but varies with $t$ as follows. 
\begin{eqnarray}
\beta=\beta(t)=\left\{ \begin{array}{ll}
1 & (nT\leq t <nT+\tau) \\
0 & (nT+\tau \leq t<(n+1)T),
\end{array} \right.
\label{eq:beta}
\end{eqnarray}
where $T$ ($T \in \mathbb{N}$) is the length of one cycle, $\tau$ ($0\leq \tau \leq T$) is the length of a period when $\beta=1$ (in the case of traffic lights, this denotes the ``green'' period), and $n \in \mathbb{N}$. In the present paper, a period with $\beta=1$ ($\beta=0$) is referred to as an (a) opening (closing) period. A schematic of $\beta(t)$ is presented in Fig. \ref{fig:exitmoshiki}. Note that particles at the right boundary, i.e., at the ($L-1$)th site, can always leave the lattice when $\beta=1$, but cannot leave the lattice when $\beta=0$. We define $\beta^*$ as the time-averaged value of $\beta$.
\begin{equation}
\beta^* =\frac{\tau}{T}
\end{equation}

\renewcommand{\figurename}{Table}
\setcounter{figure}{0}
\begin{figure}[htbp]
\begin{tabular}{cc}
\begin{minipage}[c]{0.5\linewidth}
\centering
\includegraphics[keepaspectratio, scale=0.34]{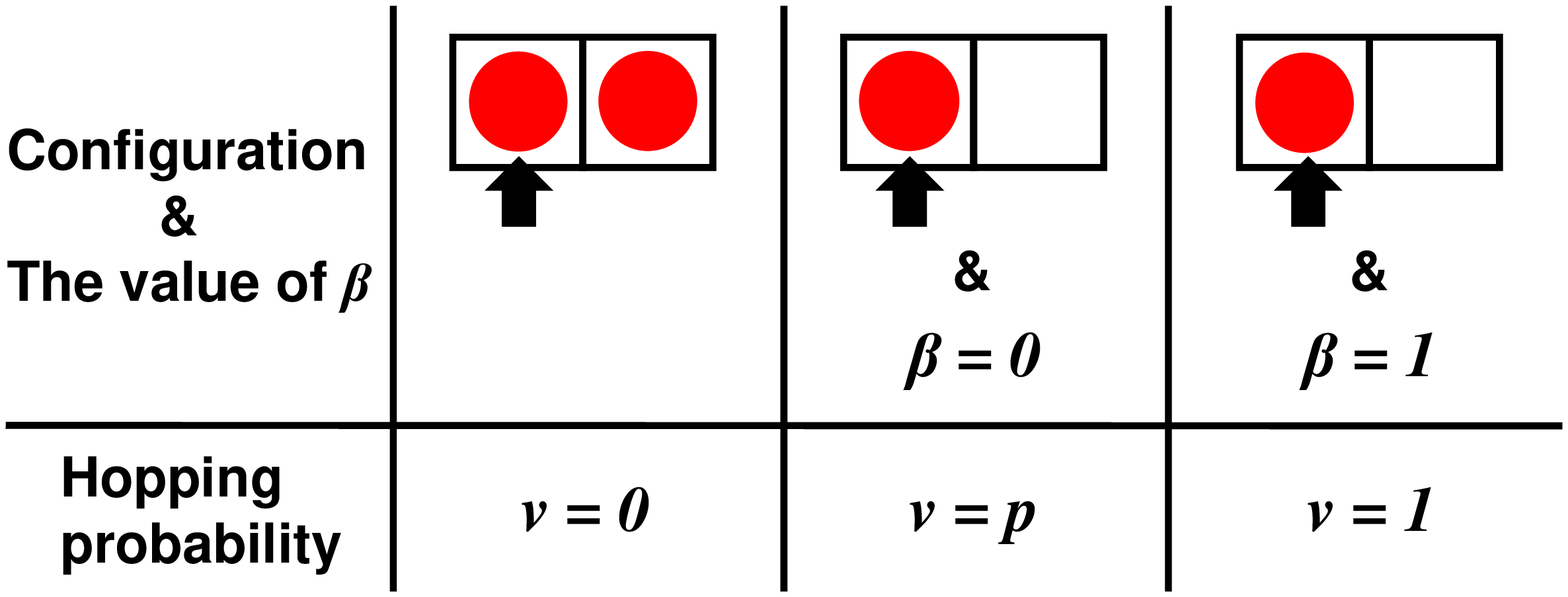}
\subcaption{}\label{fig:table1}
\end{minipage} &
\begin{minipage}[c]{0.5\linewidth}
\centering
\includegraphics[keepaspectratio, scale=0.34]{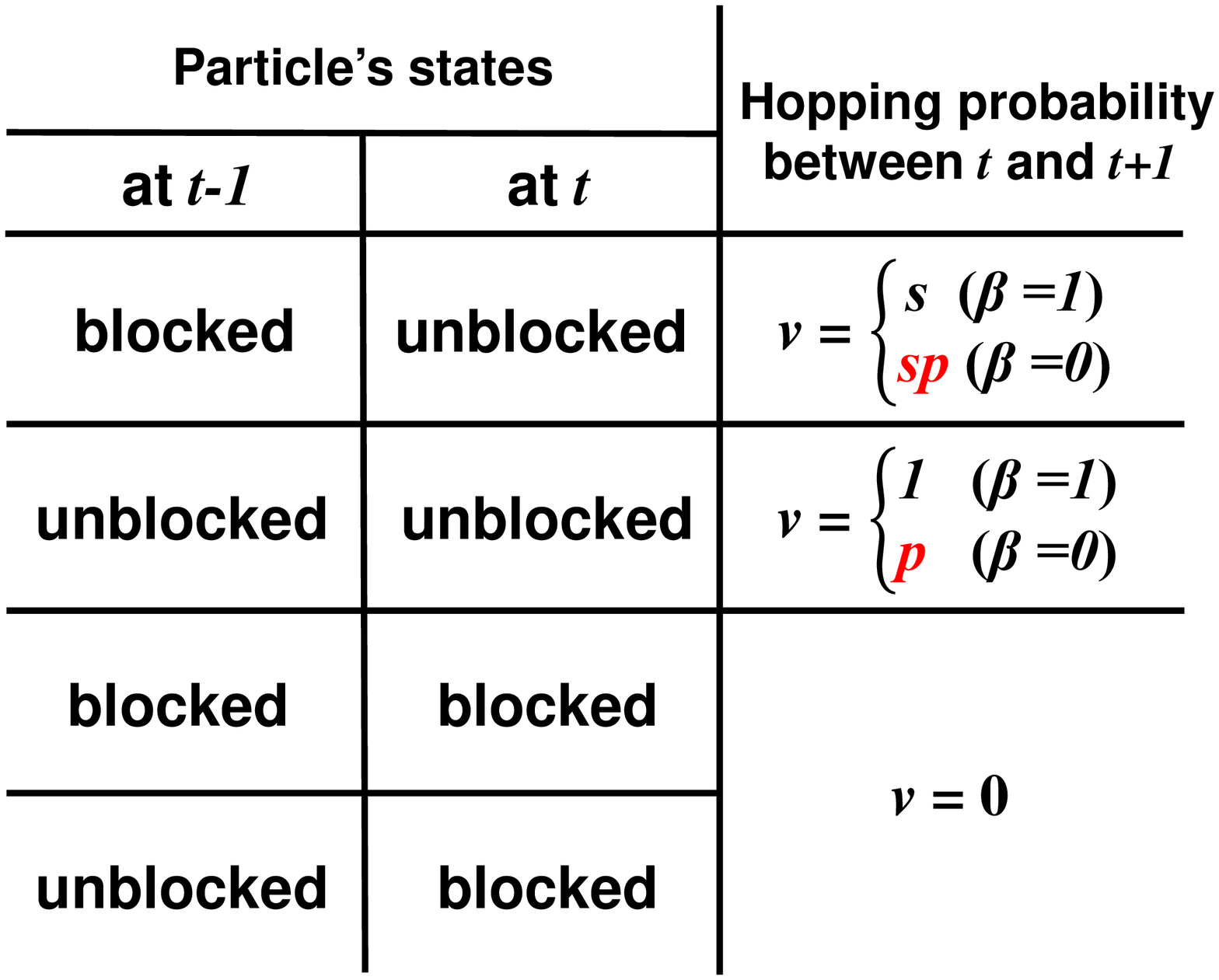}
\subcaption{}\label{fig:table2}
\end{minipage}
\end{tabular}
\caption{(a) Hopping probability of the left particle (indicated by arrows). The particle moves from left to right under the control rule. The hopping probability $v$ decreases from 1 to $p$ during a closing period ($\beta=0$). (b) Explanation of the SlS rule for a given particle. The hopping probability of the particle between time $t$ and $t+1$ is determined by the particle's state at time $t-1$ and time $t$. The labels ``blocked'' and ``unblocked'' denote the presence and absence of a particle in the right-neighboring site, respectively.}
\end{figure}

\renewcommand{\figurename}{Figure}

Next, we introduce a particle control rule. In this rule, hopping probability $v$ depends on the value of $\beta$. Particles can move no further than one site ahead (from left to right) with hopping probability $v \in \{0,p,1\}$, where $0\leq p\leq1$, as shown in Table \ref{fig:table1}. The value of $v$ is determined by the state of the exit and by the state of the particle's right-neighboring site. First, if the particle's right-neighboring site is occupied, hopping probability $v$ is set to 0. Second, if $\beta=0$ and the particle's right-neighboring site is empty, hopping probability $v$ is set to $p$. Finally, if $\beta=1$ and the particle's right-neighboring site is empty, hopping probability $v$ is set to 1. Under these rules, particles not yet involved in a jam near the exit are expected to avoid the jam. In contrast to the related works~\cite{1742-5468-2014-10-P10019, 1742-5468-2009-02-P02012, 1742-5468-2012-05-P05008, 1742-5468-2008-06-P06009, PhysRevE.87.062818}, in which hopping probability is fixed, we adopt a variable hopping law. 

We introduce a parameter $p$, representing the degree of the particles' deceleration from 1. When $p=1$, there is no control (no deceleration), whereas $0\leq p <1$ indicates that particles are decelerated during closing periods. 

Finally, all particles obey the slow-to-start (SlS) rule. The SlS rule accounts for the law of inertia and the reaction delay, realizing a more realistic model. If one particle is blocked by another particle at time $t-1$, its hopping probability between time $t$ and time $t+1$ is 0, $s$, or $sp$, depending on other conditions (see Table \ref{fig:table2}). Here $s$ is the SlS coefficient, which takes values from 0 to 1. A relatively small (large) $s$ indicates large (small) inertia and slow (fast) reaction delay of particles. If $s = 0$, particles must rest for one time step after being blocked by other particles; if $s = 1$, there is no SlS effect. 
Note that our model is generally difficult to analyze from a theoretical perspective because (i) the SlS rule increases the particle states and (ii) $\beta$ is time-inhomogeneous.

\section{Fundamental and Phase diagram without control}
In this section, we investigate and depict the fundamental diagram and phase diagram of the model when $p=1$ (indicating no control) to observe the control effects in the following sections.

The average bulk density $\overline{\rho_{\rm bulk}}$ and the average global density $\bar{\rho}$ are defined as the average number of occupancy over the space $[0.1L, 0.9L]$ and $[1, L]$ in one time step, respectively. Similarly, the average flow $Q$ is defined as the mean number of particles exiting the right boundary in one time step. As for the fundamental diagram, we observe a thousand sets of ($\overline{\rho_{\rm bulk}}$, $Q$) for every $T\times10^4$ after $T\times10^4$ steps. On the other hand, as for the phase diagram we observe ($\bar{\rho}$, $Q$) for $T\times10^6$ after $T\times10^4$ steps with each set of ($\alpha$, $\beta^*$). Note that the simulation period is fixed as $[T\times10^4, T\times10^6]$ from the next section. By evolving the lattice through $0 \leq t <T\times10^4$, we obtain $\overline{\rho_{\rm bulk}}$, $\bar{\rho}$, and $Q$ in the steady state. Finally, in both cases, $L$ is fixed as $L=200$. The results with $L=1000$ are almost the same as those with $L=200$. Thus, we choose $L=200$ to decrease calculation time.

\subsection{Fundamental diagram}

The fundamental diagram of the TASEP with the SlS rule is mainly constituted of free flow phase and jam flow phase. In the free flow phase, where every particle is not affected by the SlS rule, the average flow $Q_{\rm free}$ is equal to the average bulk density $\overline{\rho_{\rm bulk}}$.

As $\overline{\rho_{\rm bulk}}$ exceeds the critical density $\rho_{\rm cr}$, the lattice is reduced to the jam flow phase, where it is divided into two kinds of regions; a cluster region and a non-cluster region. Note that the lattice is divided into many regions, which are either a cluster region or a non-cluster region. In the cluster region, the average density is equal to 1. In the non-cluster region, the interval of particles can be two values by the SlS rule; 1 site with probability $s$ and 2 site with probability $1-s$. Therefore, the average interval of particles is reduced to $2-s$ site. As a result, the average density becomes $1/(3-s)$, which corresponds to the critical density $\rho_{\rm cr}$.
Assuming that the total length of non-cluster regions accounts for a ratio $x$ of the lattice, we obtain Eq. (\ref{eq:fd}) below;

\begin{equation}
\overline{\rho_{\rm bulk}}=\frac{1}{3-s} \times x+1 \times (1-x).
\label{eq:fd}
\end{equation}
Substituting Eq. (\ref{eq:fd}) into $Q_{\rm jam}=x/(3-s)$, we obtain

\begin{equation}
Q_{\rm jam}=\frac{1}{2-s}(1-\overline{\rho_{\rm bulk}}).
\end{equation}

Finally, even if $\overline{\rho_{\rm bulk}}$ exceeds $\rho_{\rm cr}$, metastable state sometimes appears in the case where the SlS rule never occurs, i.e., every particle maintains a gap more than or equal to one site between them. We see the metastable state until $\overline{\rho_{\rm bulk}}$ becomes 1/2, where particles maintain one-site gap between them. In the metastable state, the average flow $Q_{\rm meta}$ is equal to $\overline{\rho_{\rm bulk}}$ similar to the free flow phase.

Eventually, we can describe $Q$ as follows;

\begin{eqnarray}
Q=\left\{ \begin{array}{ll}
Q_{\rm free}=\overline{\rho_{\rm bulk}} & \rm{for} \ \overline{\rho_{\rm bulk}}<\frac{1}{3-s} \\
Q_{\rm jam}=\frac{1}{2-s}(1-\overline{\rho_{\rm bulk}}) & \rm{for} \ \overline{\rho_{\rm bulk}}>\frac{1}{3-s}\\
Q_{\rm meta}=\overline{\rho_{\rm bulk}} & \rm{for} \ \frac{1}{3-s}<\overline{\rho_{\rm bulk}}<\frac{1}{2}.
\end{array} \right.
\label{eq:fd2}
\end{eqnarray}

Figure \ref{fig:fd} plots the simulated (red dots) and theoretical (green line) with (a) $s=0$ and (b) $s=0.5$. In both figures, the simulated values favorably agree with the theoretical line, Eq. (\ref{eq:fd2}). Note that some of the simulated values which take values in $Q=(Q_{\rm jam}, Q_{\rm meta}]$ can be obtained only with the value of $\beta^*$ very close to 1.

\setcounter{figure}{2}
\begin{figure}[h]
\begin{tabular}{cc}
\begin{minipage}[t]{0.5\linewidth}
\centering
\includegraphics[keepaspectratio, height=5.5cm]{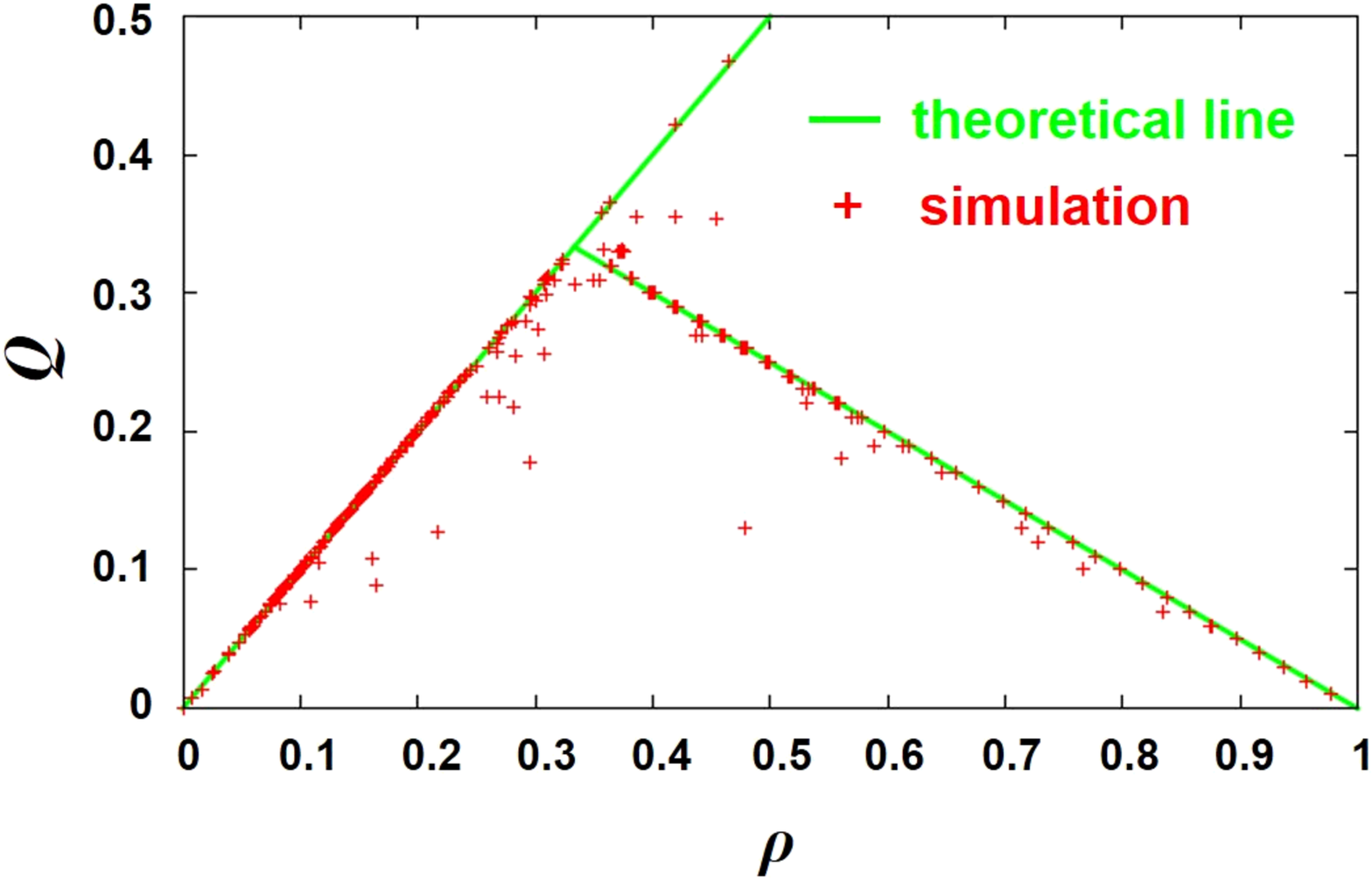}
\subcaption{}\label{fig:fds0}
\end{minipage} &
\begin{minipage}[t]{0.5\linewidth}
\centering
\includegraphics[keepaspectratio, height=5.5cm]{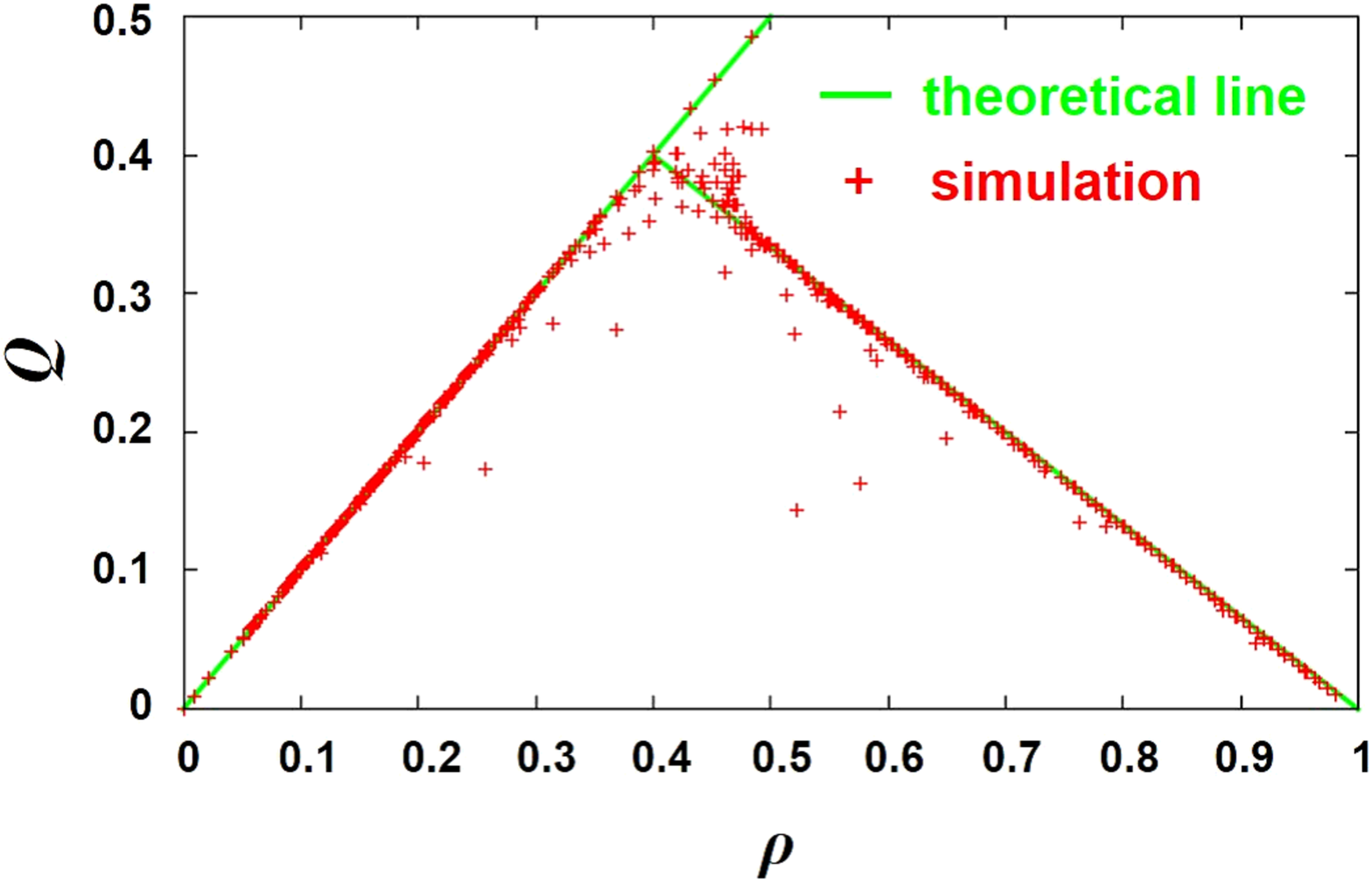}
\subcaption{}\label{fig:fds0.5}
\end{minipage}
\end{tabular}
\caption{Fundamental diagram for (a) $s=0$ and (b) $s=0.5$, plotting the simulated (red dots) and theoretical (green line) values. The other parameters are set as $L=200$, $T=100$, and $p=1$. The simulated values favorably agree with the theoretical line in the both figures. The theoretical results are obtained by Eq. (\ref{eq:fd2}).}
\label{fig:fd}
\end{figure}

\subsection{Phase diagram}
\begin{figure}[h]
\begin{tabular}{cc}
\begin{minipage}[t]{0.5\linewidth}
\centering
\includegraphics[keepaspectratio, height=7cm]{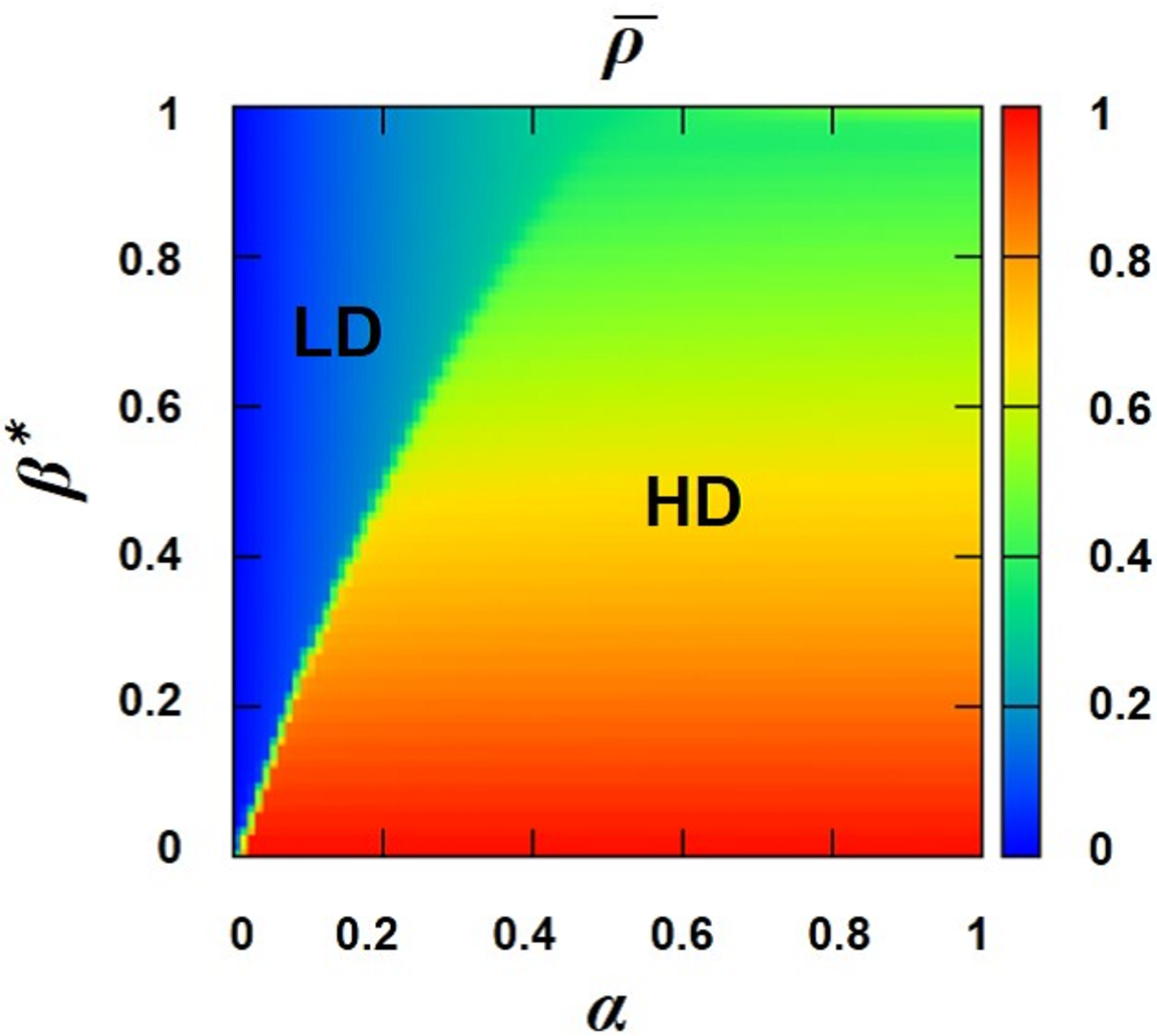}
\subcaption{}\label{fig:phasea}
\end{minipage} &
\begin{minipage}[t]{0.5\linewidth}
\centering
\includegraphics[keepaspectratio, height=7cm]{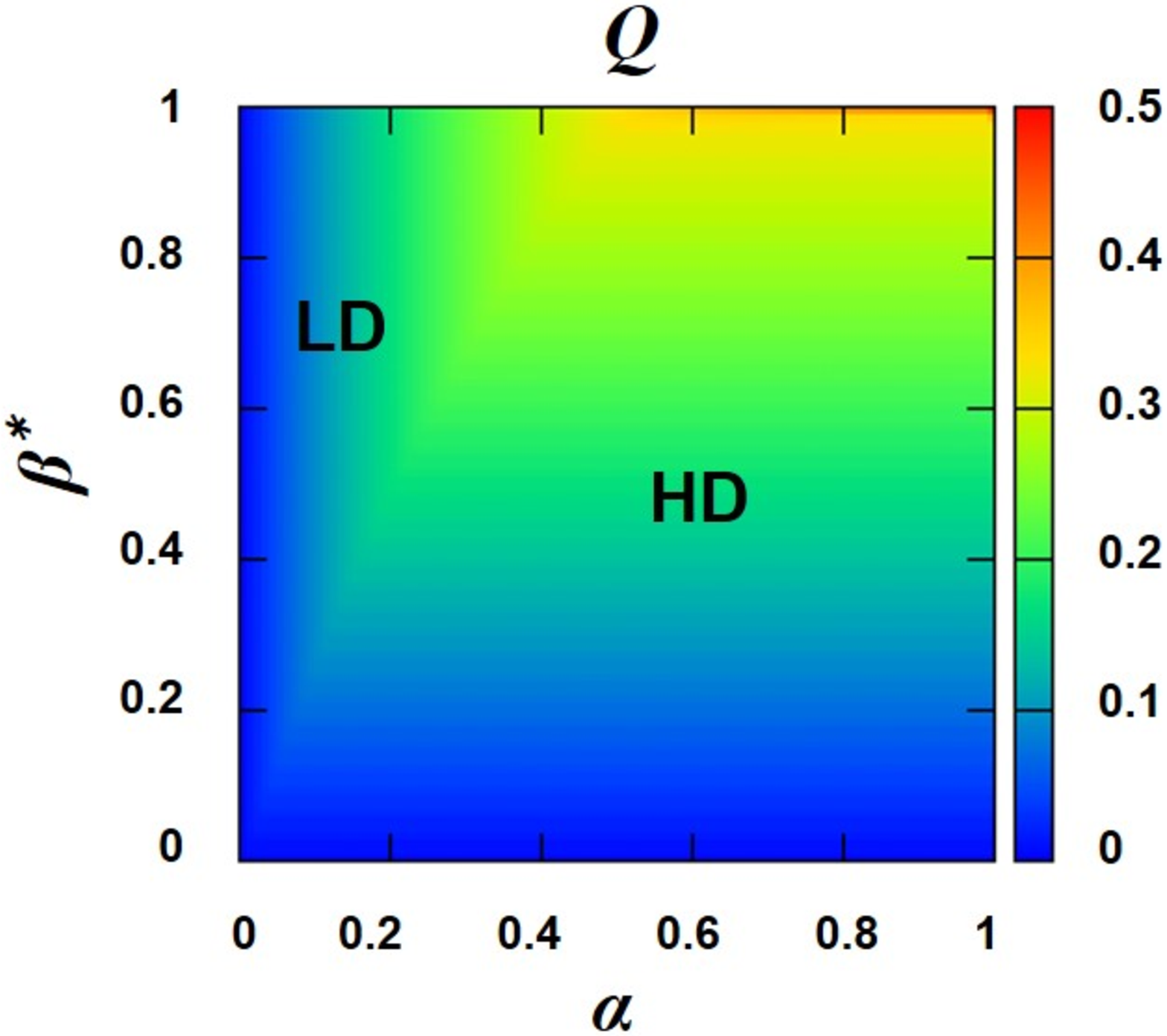}
\subcaption{}\label{fig:phaseb}
\end{minipage}
\end{tabular}
\caption{Phase diagram without the control. Color bar indicates the values of $\bar{\rho}$ in (a) ($Q$ in (b)). The other parameters are set as $L=200$, $T=100$, $p=1$, and $s=0$.}
\label{fig:phase}
\end{figure}

In this subsection, we discuss and depict the phase diagram of the model.
The results of the numerical simulation are shown in Fig. \ref{fig:phase} by plotting the values of $\bar{\rho}$ (Fig. \ref{fig:phasea}) and $Q$ (Fig. \ref{fig:phaseb}). The other parameters are set as $T=100$ and $s=0$.

Figure \ref{fig:phase} reveals three phases of different densities: a high density (HD) phase, a low density (LD) phase, and a metastable state. No maximum current (MC) phase exists, as the maximal hopping probability is set to 1. 
We stress that the representations of "LD" and "HD" is determined only by the set of ($\alpha$, $\beta^*$).

\begin{figure}[htbp]
\begin{center}
\includegraphics[width=6cm,clip]{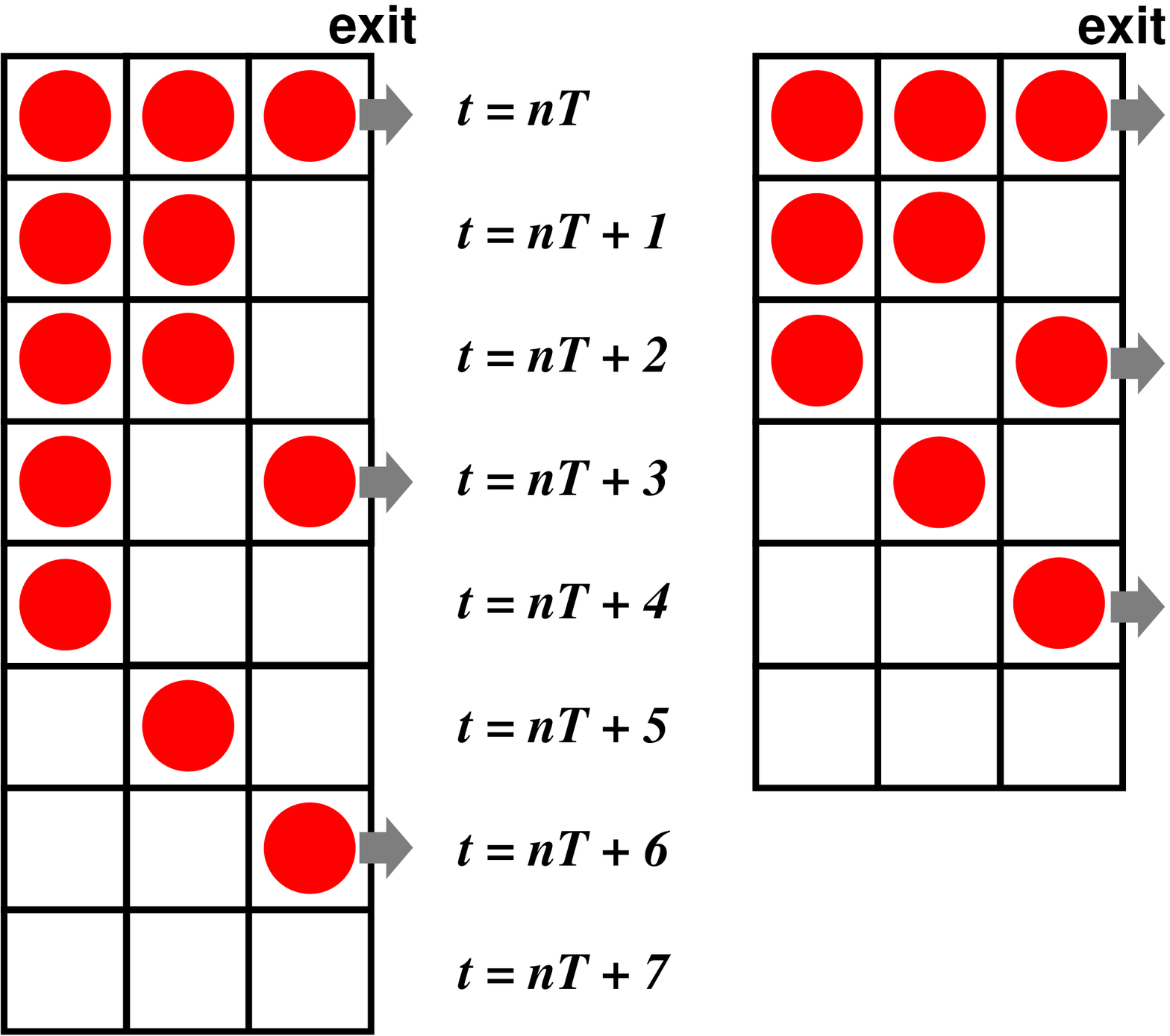}
\caption{Schematic of three sites near the exit in the HD phase. The time series is $t=nT$ to $t=nT+7$ (left: SlS rule, right: no SlS rule). The right figure is shown for comparison. In the left (right) panels, each particle exits the lattice in every 3 (2) time steps in the left (right) figure. In this schematic, three particles accumulate in front of the exit during the closing period. Note that the parameter $s$ is set as $s=0$ in the left panel.}
\label{fig:deguti}
\end{center}
\end{figure}

In the HD phase, $Q$ is governed by $\beta^*$, the time-averaged value of $\beta$, which is defined in Eq. (\ref{eq:beta}).
In this case, a jam is often formed in front of the exit. Therefore, all particles which attempt to leave the lattice are influenced by the SlS rule. As shown in Fig. \ref{fig:deguti} (left), one particle exits the lattice in every 3 time steps. Therefore, assuming a sufficient number of particles occupy the front of the exit, the average number of particles leaving the lattice is $\lceil \tau/3 \rceil$ ($=\lceil \beta^*T/3 \rceil$) per cycle. Eventually, $Q$ settles to $\lceil \beta^*T/3 \rceil /T$. 
Note that $\lceil x \rceil$ denotes the smallest integer greater than or equal to $x$.

On the other hand, the SlS rule has little effect on the flow in the LD phase, because the opening period is long enough to let all particles accumulating in front of the exit evacuate.
Therefore, in this phase, $Q$ is governed by the input probability of particles from the left boundary. Consequently, $Q$ is given by $\alpha/(1+\alpha)$, which describes the particle flow in the original TASEP with parallel updating~\cite{PhysRevE.59.4899}.

At the boundary between the LD and HD phases, the flows must match~\cite{PhysRevLett.86.2498}. Therefore, we have:
\begin{equation}
\left\lceil \frac{\beta^*T}{3} \right\rceil\frac{1}{T}=\frac{\alpha}{1+\alpha}.
\label{eq:boundary}
\end{equation}
From Eq. (\ref{eq:boundary}), the boundary between the HD and LD phases is described by $\beta^* = 3\alpha/(1+\alpha)$ when $\tau$ $(=\beta^*T)$ is a multiple of 3. On the other hand, when $\tau=3m-1$, the left side of Eq. (\ref{eq:boundary}) becomes $(\tau+1)/3T$, where $m \in \mathbb{N}$. Similarly, when $\tau=3m-2$, the left side of Eq. (\ref{eq:boundary}) becomes $(\tau+2)/3T$. 

Finally, when $\alpha$ exceeds $1/2$ under the condition $\beta^*=1$, particles never stop at the exit and are never affected by the SlS rule. Therefore, we observe metastable state, where the flow is equal to $Q=\alpha/(1+\alpha)$ exceeding the maximal current 1/3.


\begin{figure}[h]
\begin{tabular}{cc}
\begin{minipage}[t]{0.5\linewidth}
\centering
\includegraphics[keepaspectratio, height=5cm]{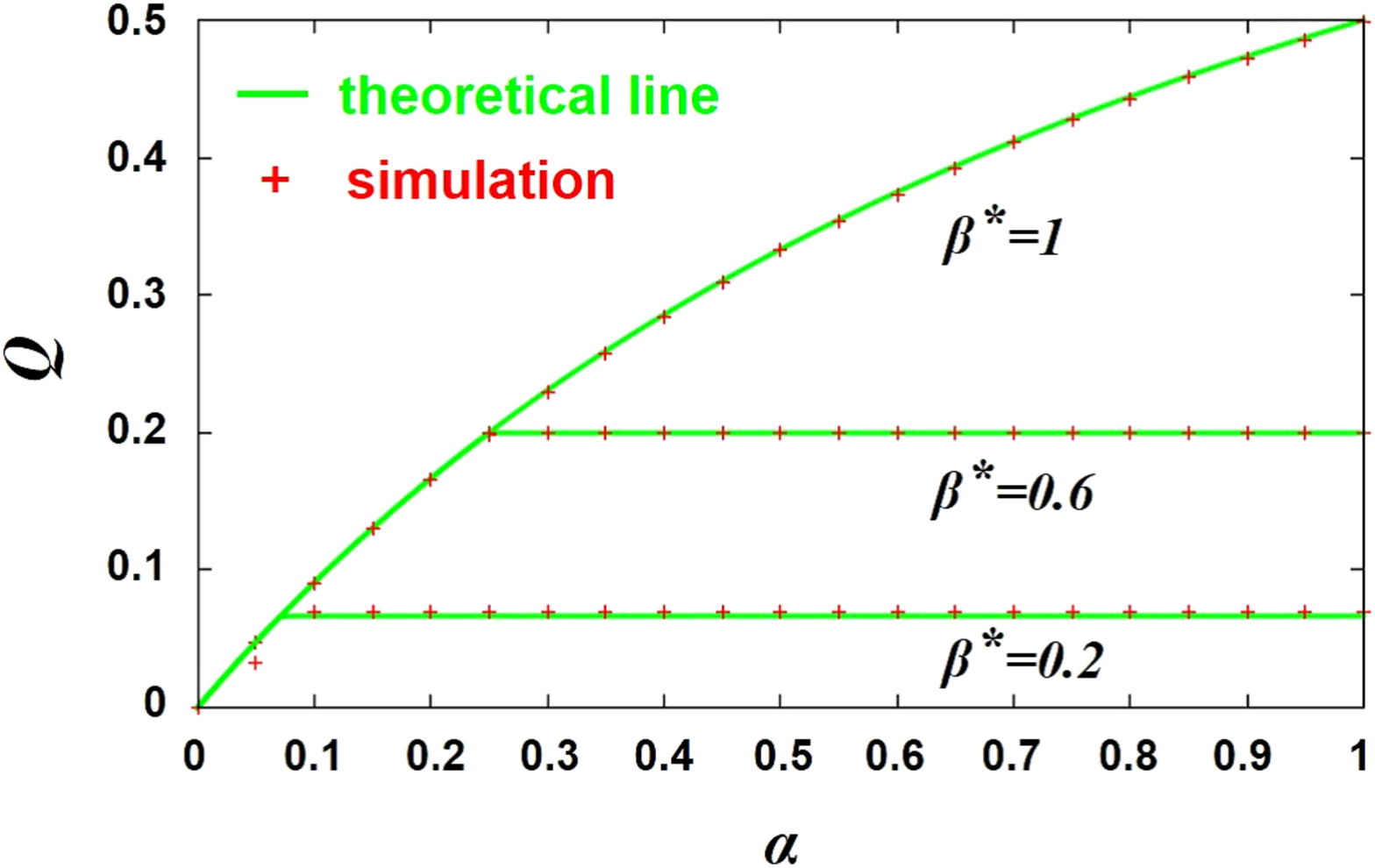}
\subcaption{}\label{fig:comment1-2-1}
\end{minipage} &
\begin{minipage}[t]{0.5\linewidth}
\centering
{\includegraphics[keepaspectratio, height=5cm]{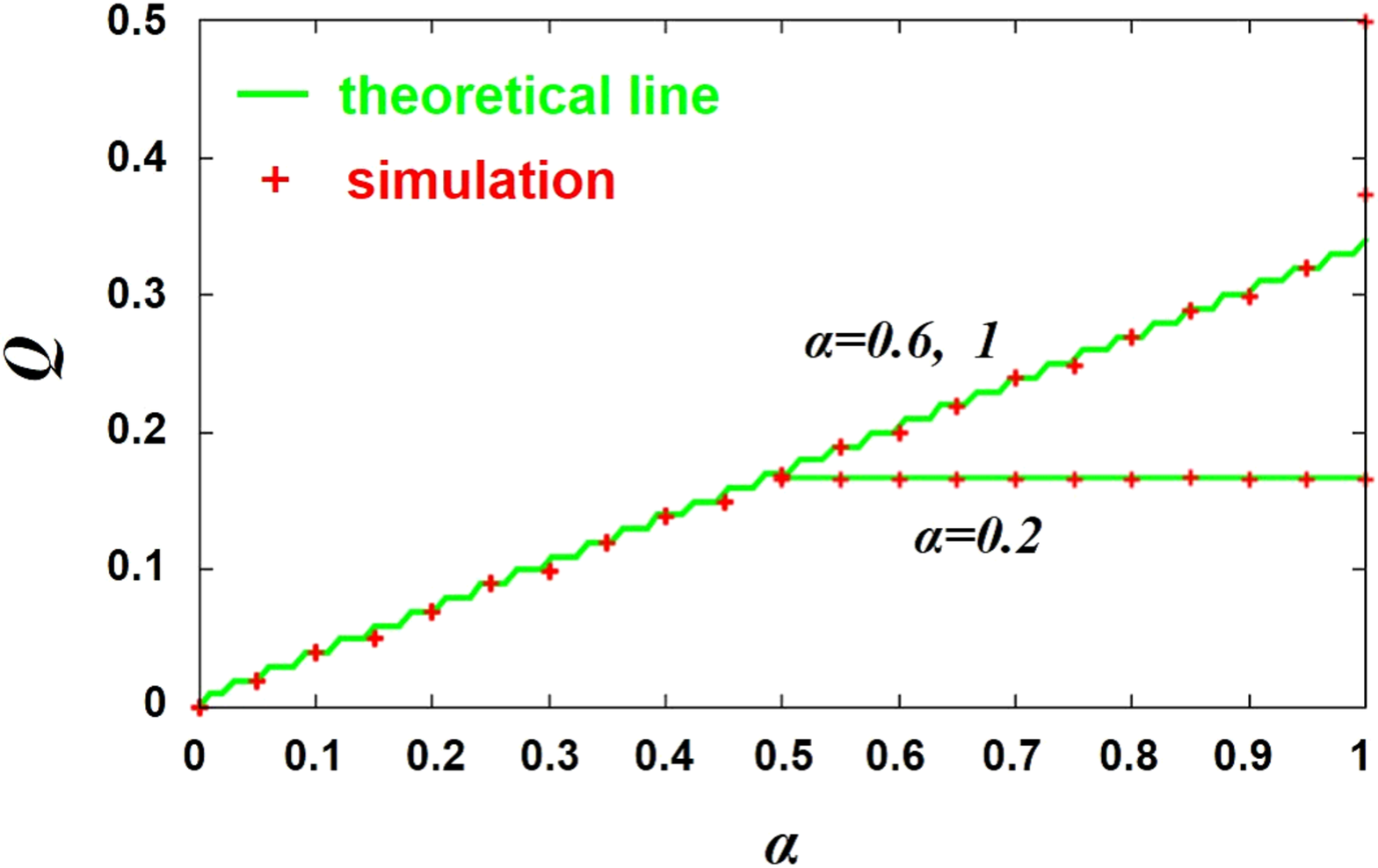}}
\subcaption{}\label{fig:comment1-2-2}
\end{minipage} 
\end{tabular}
\caption{(a) Simulated (red dots) and theoretical (green line) of $Q$ as a function of $\alpha$ for various $\beta^* \in \{0.2, 0.6, 1\}$. (b) Simulated (red dots) and theoretical (green line) of $Q$ as a function of $\beta^*$ for various $\alpha \in \{0.2, 0.6, 1\}$. The other parameters are set as $L=200$, $T=100$, $p=1$, and $s=0$.}
\label{fig:phasedash}
\end{figure}


Note that in the case of $s=1$ (no SlS effect) we can derive the boundary between HD and LD with similar discussion (see Fig. \ref{fig:deguti}) as follows:

\begin{equation}
\left\lceil \frac{\beta^*T}{2} \right\rceil\frac{1}{T}=\frac{\alpha}{1+\alpha}.
\label{eq:boundarydash}
\end{equation}
The boundary between the LD and HD phases moves from Eq. (\ref{eq:boundarydash}) to Eq. (\ref{eq:boundary}) as $s$ changes from $s=1$ to $s=0$.

Figure \ref{fig:phasedash} compares the simulated values and the theoretical line of $Q$ with (a) a fixed $\beta^*$=0.6 or (b) a fixed $\alpha$=0.2. In Fig. \ref{fig:phasedash}, the simulated values favorably agree with the theoretical line. We also confirm the metastable state whose flow is equal to $Q=\alpha/(1+\alpha)$ when $\beta^*=1$ in Fig. \ref{fig:comment1-2-1}. Note that in Fig. \ref{fig:comment1-2-2} the simulated values for $\beta^*=1$ seem to be deviated from the theoretical line because those cases are in the metastable state.

\section{Investigation of flow improvement with control}
In this section, we investigate whether the average flow $Q$ is improved by our control, where we vary the values of $p$ according to the state of a bottleneck. $Q$ is mainly a function of $p$; that is, we write $Q(p)$. 
We examine how $Q(p)$ changes between the cases of $p=1$ and $0<p<1$. Note that when $p=1$, particles never decelerate (i.e., no control), whereas when $0<p<1$, particles decelerate during a closing period (indicating that the control works).
Hereafter, we set $\beta^*=0.6$ as a representative because we can simplify Eq. (\ref{eq:boundary}) and 
observe the effect of the control in the phases which would be HD (LD) in the absence of the control by changing $\alpha$.
We discuss the simulation results by changing $\beta^* \ (\tau)$ in \ref{sec:appendixt}. We also review the dependence of the effect of the control in \ref{sec:appendixs}.

\subsection{Improvements of $Q(p)$ for various $\alpha$}
\label{sec:a}

In this subsection, we investigate whether $Q(0<p<1)$ is improved over the case of $Q(1)$ for various $\alpha$. Here, we set $T=20$ and $s=0$ to investigate the cases where we have the largest SlS effect. 

\begin{figure}[htbp]
\begin{center}
\includegraphics[width=9cm,clip]{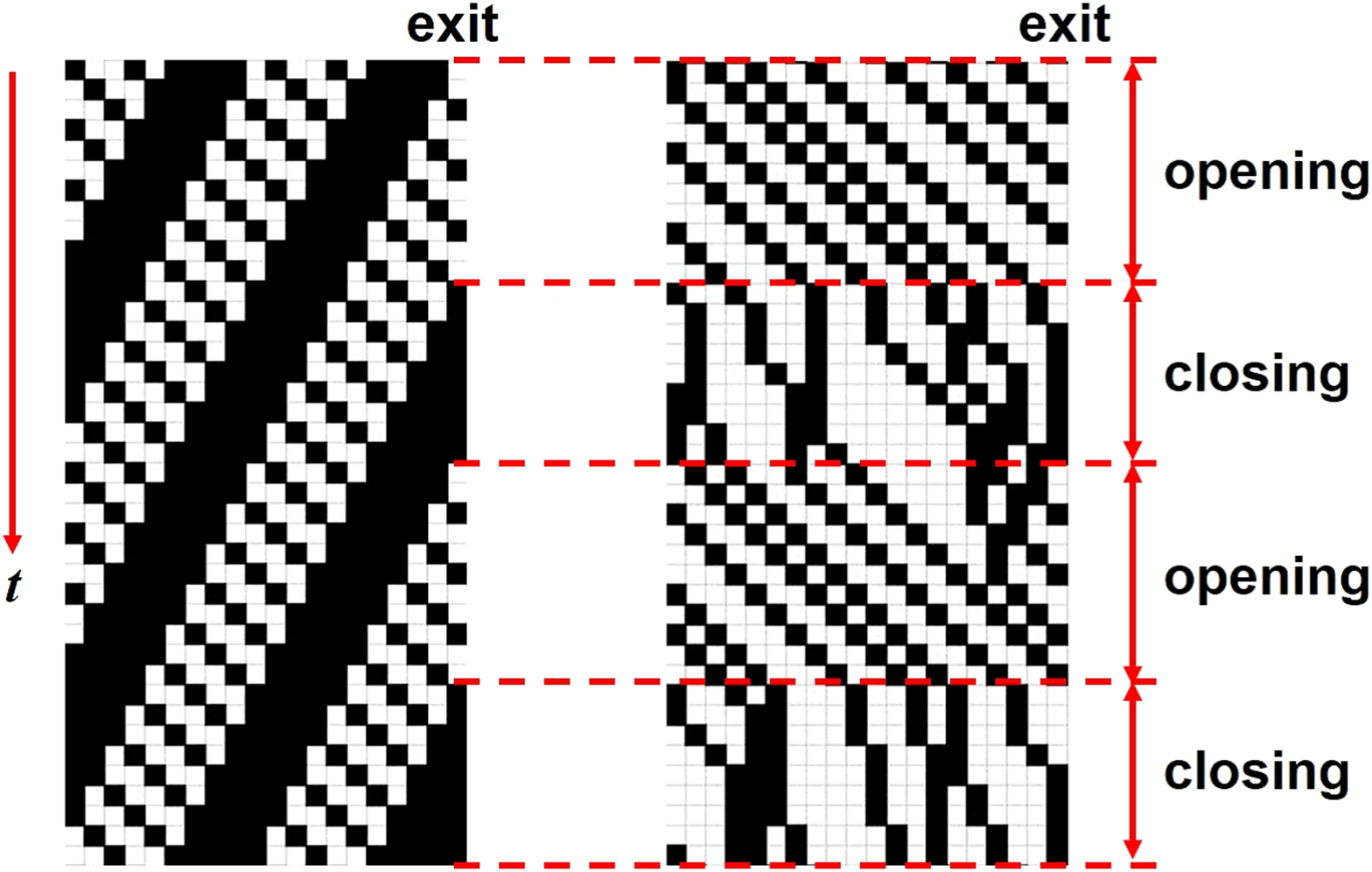}
\caption{Space-time diagrams of the model. The left and right panels show the results of the uncontrolled ($p=1$) and controlled ($p=0.3$) cases, respectively. Black (white) squares indicate sites of $n_i(t)=1$ ($n_i(t)=0$). We extract only 20 sites from the right boundary per two cycles (40 time steps) in the steady state with $\alpha=1$, $\beta^*=0.6$, and $T=20$.}
\label{fig:jikuzu}
\end{center}
\end{figure}

Figure \ref{fig:jikuzu} shows space-time diagrams of the model for $p=1$ and $p=0.3$. Under these conditions, only 4 particles per open period can leave the lattice in the uncontrolled case (the left panel). However, in the controlled case (the right panel), 5 particles leave the lattice during the second open period, because the control mitigates jamming near the exit.

Next, we vary the values of $\alpha$ and investigate the improvements of $Q(p)$ to the model. We define $a(p)$ as the ratio of the change of $Q(p)$ from $Q(1)$; that is, 
\begin{equation}
a(p)=\frac{Q(p)-Q(1)}{Q(1)}.
\end{equation} 
From the definition of $a(p)$, the cases of $a(p)>0$ ($a(p)<0$) indicate that the control is effective (detrimental).

Referring to the last section, the uncontrolled flow is $Q(1)=\alpha/(1+\alpha)$ in the LD, and $Q(1)=\beta^*/3$ in the HD. We observe the improvement or deterioration of $Q(0<p<1)$ from $Q(1)$ for $\alpha \in \{0.2, 0.4, 1\}$. We select these values of $\alpha$ to investigate the LD case, the case near the boundary and the HD case without the control. The conditions $(\alpha, \beta^*) \in \{(0.2, 0.6), (0.4, 0.6), (1, 0.6)\}$ correspond to the LD, HD, and HD, respectively, because the cases $\alpha<\beta^*/(3-\beta^*)$ and $\alpha>\beta^*/(3-\beta^*)$ represent the LD and HD phases when $p=1$. For $(\alpha, \beta^*) \in \{(0.2, 0.6), (0.4, 0.6), (1, 0.6)\}$, $Q(1)$ is equal to 1/6, 1/5, and 1/5, respectively. 

\begin{figure}[h]
\begin{center}
\includegraphics[width=9cm,clip]{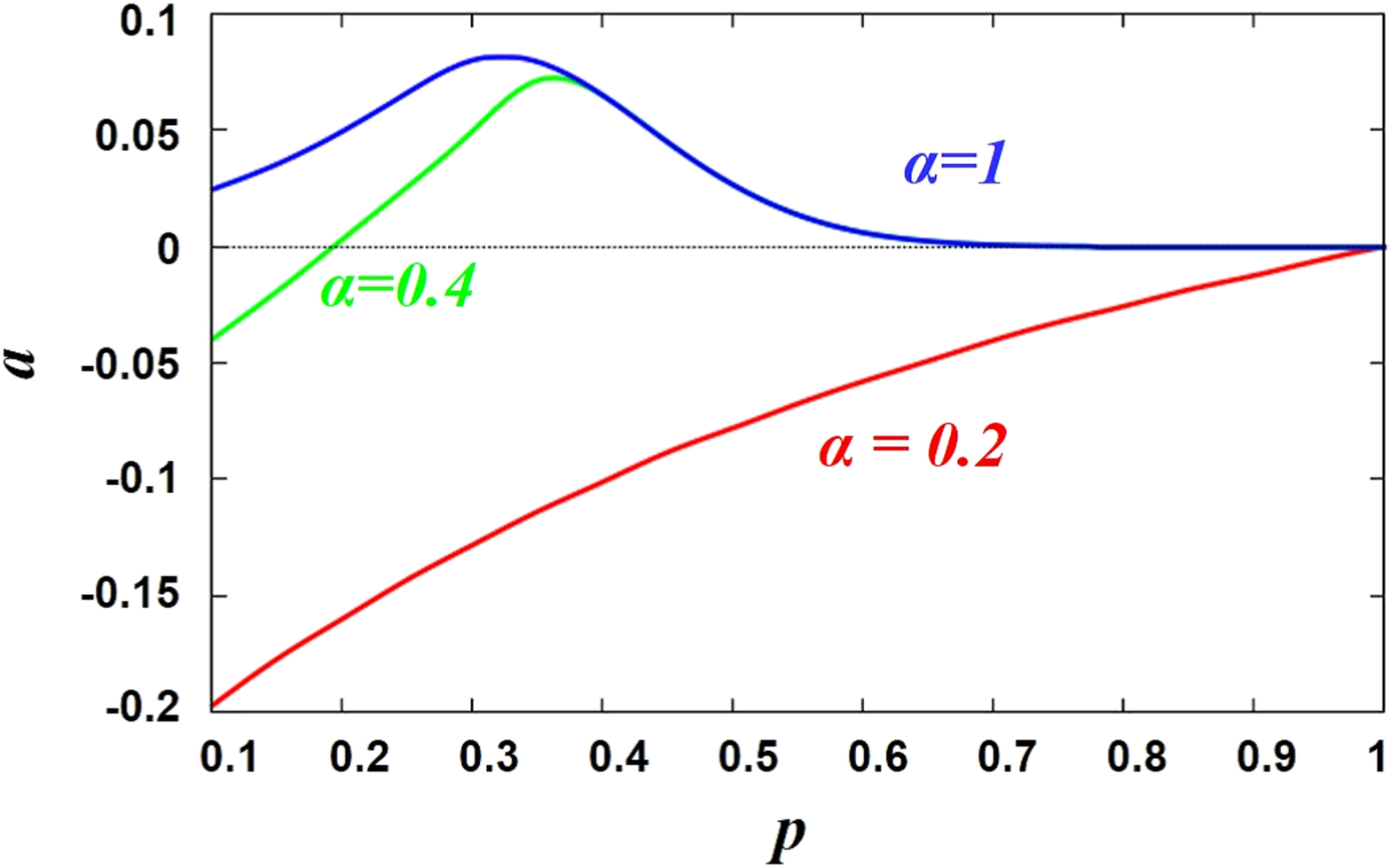}
\caption{Simulated values of $a(p)$ as a function of $p$ for various $\alpha$ $\in$ \{$0.2$(red), $0.4$(green), $1$(blue)\}. The other parameters are set as $L=200$, $\beta^*=0.6$, $T=20$, and $s=0$.}
\label{fig:kaizen1}
\end{center}
\end{figure}

The simulated values of $a(p)$ are plotted in Fig. \ref{fig:kaizen1}.
We first observe that in the case $\alpha=0.2$, $a(p)$ monotonically increases as $p$ approaches 1, indicating that the control is detrimental rather than effective. 
On the other hand, in the cases $\alpha=$0.4 and 1, the control is effective at appropriate values of $p$, and there exists an optimal $p$, $p_{\rm opt}$, that maximizes $a(p)$. 
When $p$ is too small, however, $a(p)$ decreases because the gaps between particles become excessively  long and lead to a decrease in the flow.
We have also confirmed that $a$ never exceeds 0 with Woelki's method (controlling input rate according to the lattice density)~\cite{PhysRevE.87.062818} in our model 
(see \ref{sec:appendixw} for more details)
.

Since the scope of this study is to investigate the effect of our control, we hereafter set $\alpha=1$, with which the most remarkable flow improvement is observed. 

Note that under the control, $Q(p)$ can take $\beta^*/2$ (=0.3) as the maximum value and $\beta^*/3$ (=0.2) as the minimum value in the HD phase; that is, the maximal $a(p)$ is theoretically 0.5. Therefore, we consider that $a(p)\approx0.1$ with $T=20$ achieved by our control is not a subtle increase but an effective improvement.

\begin{figure}[htbp]
\begin{center}
\includegraphics[width=9cm,clip]{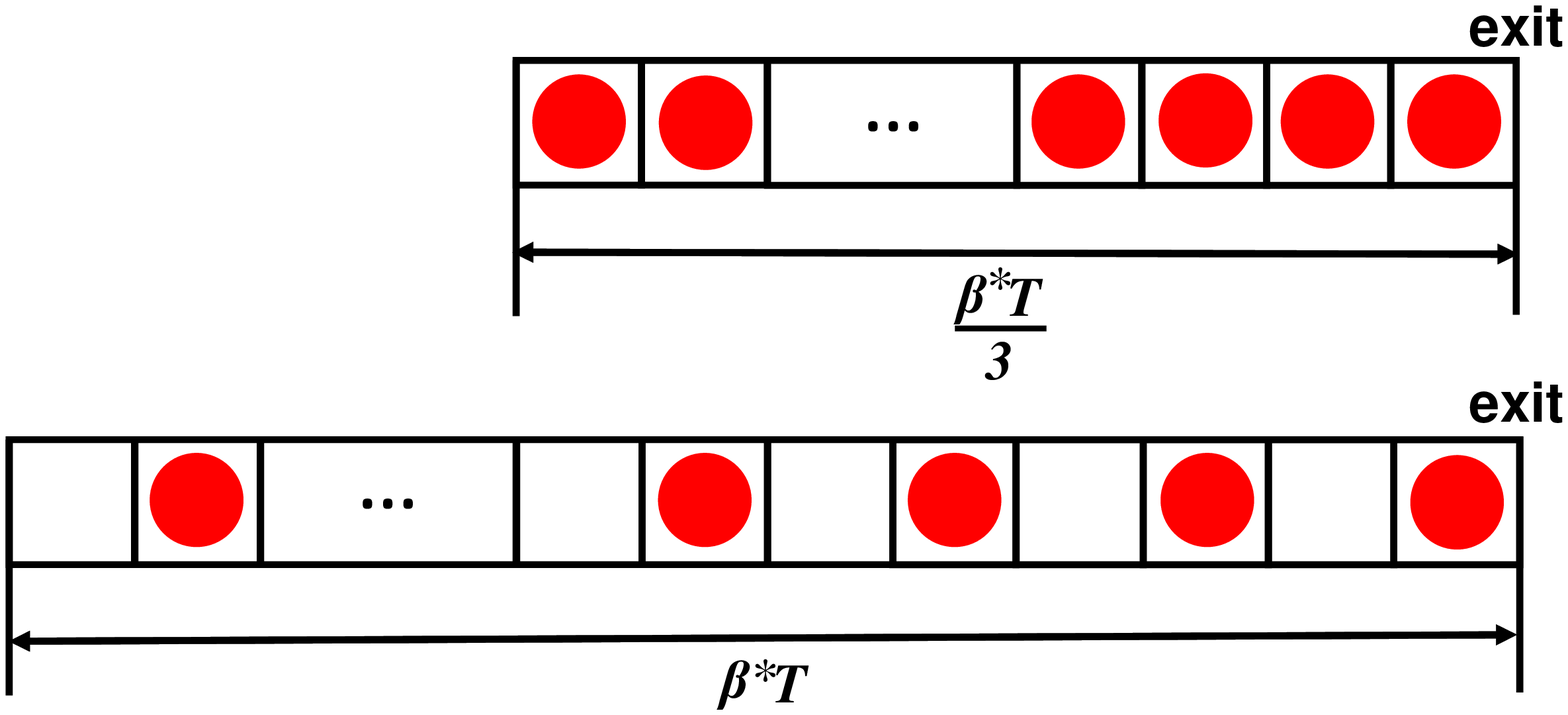}
\caption{Configurations of particles in the HD phase. The upper panel illustrates the uncontrolled case ($Q(p)=\beta^*/3=0.2$); the lower panel illustrates the ideal case ($Q(p)=\beta^*/2=0.3$). In these examples, $\beta^*T(=\tau)$ is a multiple of 6.}
\label{fig:configuration}
\end{center}
\end{figure}

Figure \ref{fig:configuration} shows the configurations in front of the exit at $t=nT$, when $\beta$ is switched from 0 to 1. The upper (lower) panel illustrates the case $Q(p)=\beta^*/3 $ (=0.2) ($Q(p)=\beta^*/2$ (=0.3)). When $\alpha=1$ and $p=0$, $Q(p)$ becomes $\beta^*/2$ (=0.3); that is, $a(p)$ equals 0.5, because every particle maintains a one-site gap. 
However, $Q(p)$ plummets as $p$ increases marginally from 0 because some particles move forward and are blocked by their leading particles. This results in the occurrence of the SlS rule, which is propagated through the lattice.
Eventually, we can regard the state with $p=0$ as a sort of metastable state. Such a special case is out of our scope, so that we search the range from $p=0.1$ to $p=1$ in Fig. \ref{fig:kaizen1} and \ref{fig:kaizen2}.

\subsection{Improvements of $Q(p)$ for various $T$}
\label{sec:b}
\begin{figure}[h]
\begin{center}
\includegraphics[width=9cm,clip]{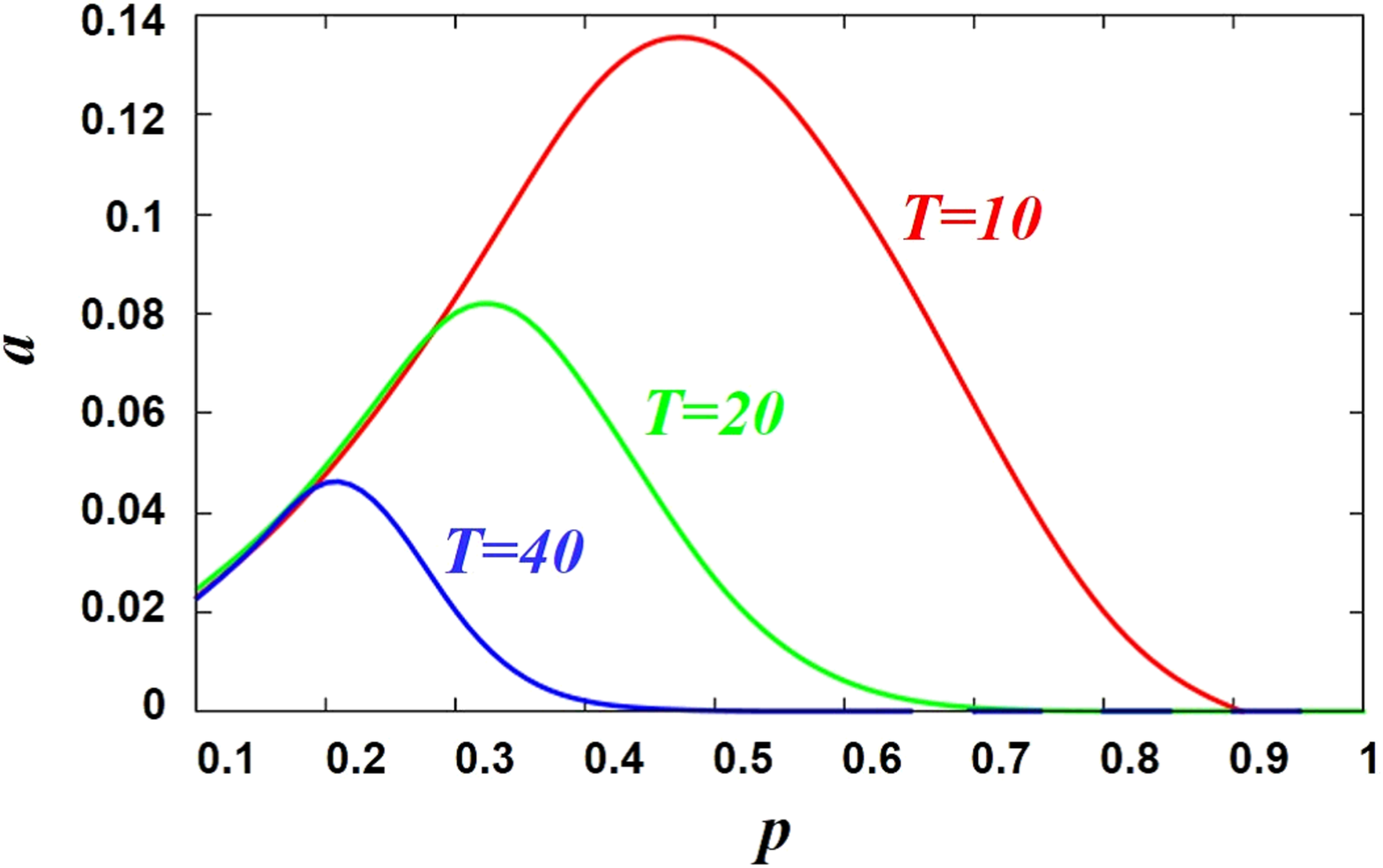}
\caption{Simulated values of $a(p)$ as a function of $p$ for various $T\in$\{10 (red), 20 (green), 40 (blue)\}. The other parameters are set as $L=200$, $\alpha=1$, $\beta^*=0.6$, and $s=0$.}
\label{fig:kaizen2}
\end{center}
\end{figure}

Next, we investigate the effect of $T$ on $a(p)$ with $s=0$ for constant $\beta^*=0.6$. Figure \ref{fig:kaizen2} plots the simulated values of $a(p)$ as a function of $p$ for various $T \in \{10, 20, 40\}$.

Two phenomena are observed in Fig. \ref{fig:kaizen2}. 
First, the maximum $a(p)$ increases as $T$ decreases because particles less readily accumulate in front of the exit with a shorter closing period. 
Second, $p_{\rm opt}$ decreases as $T$ increases. As a closing period grows, larger gaps between particles produced with smaller values of $p$ are required in order to remove jams. Eventually, the relatively long jams with a relatively long $T$ become difficult to absorb; that is, $a(p)$ approaches 0.

\begin{figure}[h]
\begin{center}
\includegraphics[width=6cm,clip]{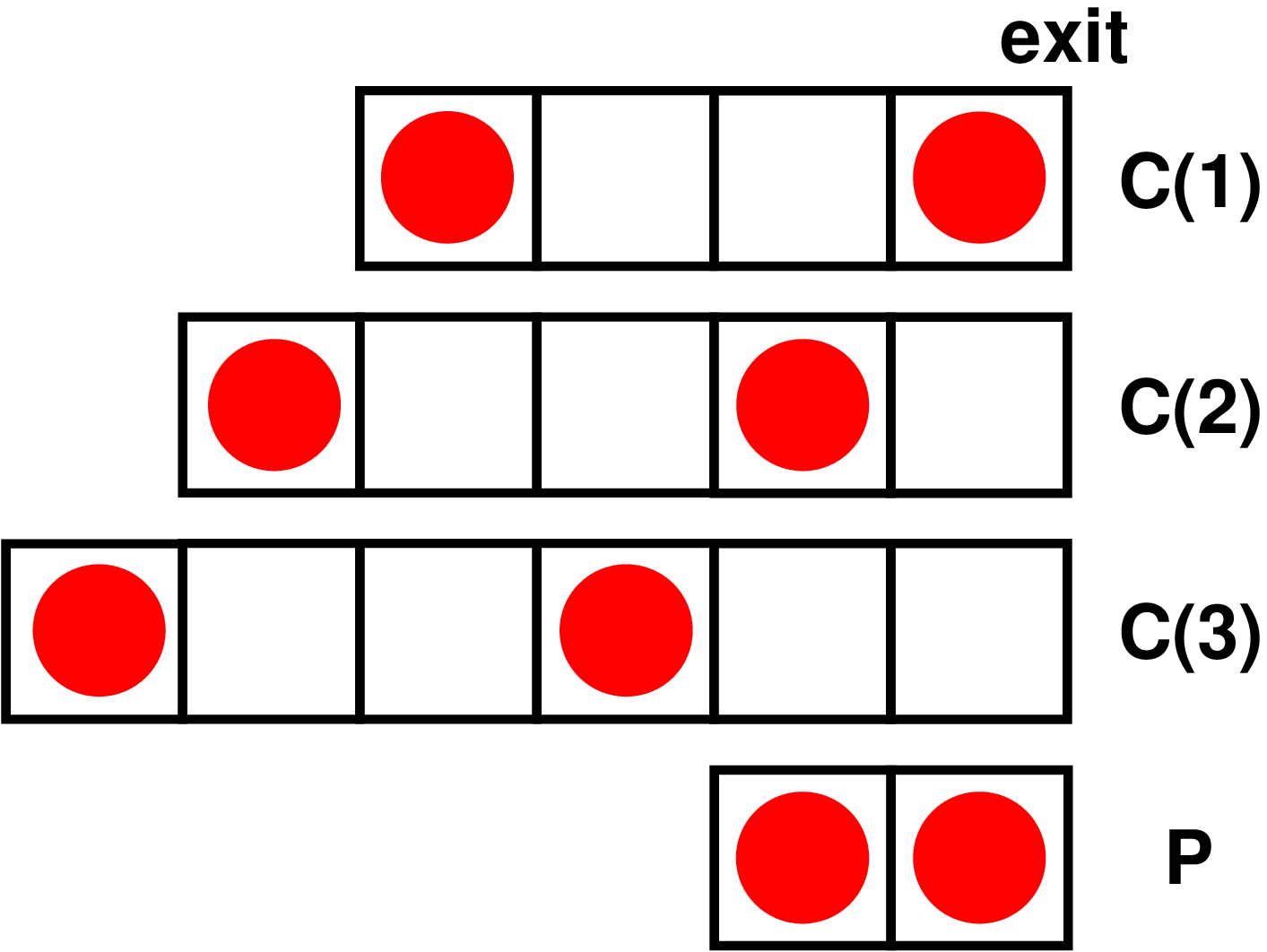}
\caption{Three different particle configurations, labeled C(1), C(2), and C(3) at $t=nT+\tau$, and the configuration of the starting point of a jam, labeled P.} 
\label{fig:exit}
\end{center}
\end{figure}

We also discuss the relation between $T$ and $p_{\rm opt}$ in numerical simulations and theoretical analyses.
In the HD phase, we assume that particles relatively near the exit are always separated by a two-site gap, because almost all of the particles are influenced by the SlS rule until they approach the exit. Under this assumption, the configuration of two particles near the exit at $t=nT+\tau$, when $\beta$ is switched from 1 to 0, can take three different states (labeled C(1), C(2), and C(3) in Fig. \ref{fig:exit}) with equal probability. Neglecting the occasions when the following particle catches up with the leading particle between $t=nT+\tau$ and $t=(n+1)T$ in C(2) and C(3), the states C(1), C(2), and C(3) reach the state P in times $2/p$, $3/p$, and $4/p$ on average, respectively. Note that if the state P is reached during a closing period, $Q$ approaches $\beta^*/3 (=0.2)$ because the particles following the two particles depicted in Fig. \ref{fig:exit} are also influenced by the SlS rule. Therefore, in the ideal case, the following particle will on average reach the second site from the exit at $t=(n+1)T$, when $\beta$ is switched from 0 to 1 and the leading particle leaves the lattice. In this case, the state P can be avoided with the minimum gap. Formulating this condition, we obtain Eq. (\ref{eq:approximate}) below;

\begin{equation}
T-\tau=(1-\beta^*)T=\frac{2}{p} \times \frac{1}{3} +\frac{3}{p} \times \frac{1}{3} +\frac{4}{p} \times \frac{1}{3}.
\label{eq:approximate}
\end{equation}
Substituting $p=p_{\rm opt}$ into Eq. (\ref{eq:approximate}), we obtain the simplified expression:

\begin{equation}
p_{\rm opt}=\min\left\{\frac{3}{(1-\beta^*)T}, \ 1\right\}.
\label{eq:popt1}
\end{equation}

\begin{figure}[h]
\begin{center}
\includegraphics[width=9cm,clip]{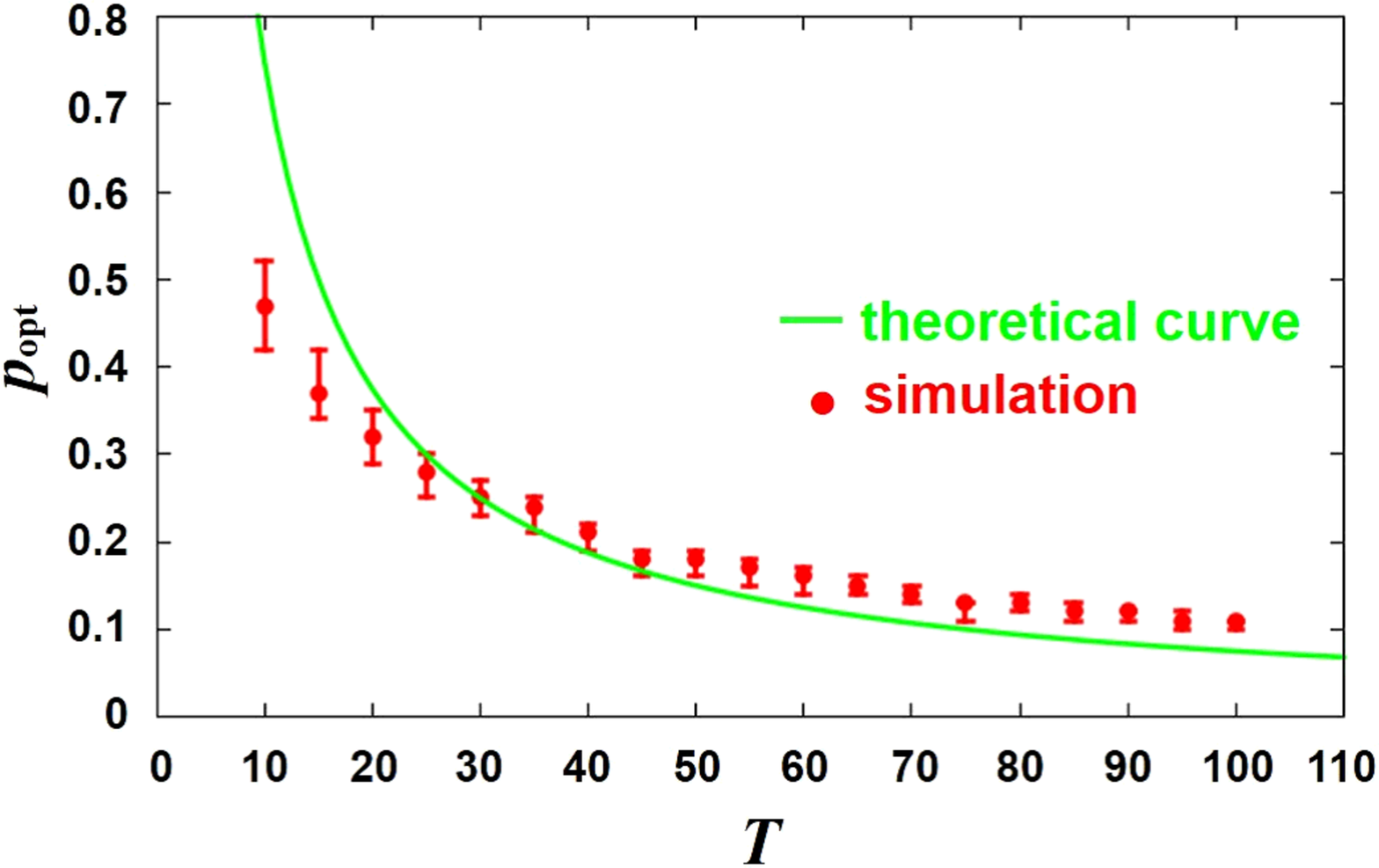}
\caption{Simulated (red dots) and theoretical (green curve) values of $p_{\rm opt}$ as a function of $T$. The red bars show the range of $p$ which achieves more than 95\% of the improvement with $p_{\rm opt}$. The other parameters are set as $L=200$, $\alpha=1$, $\beta^*=0.6$, and $s=0$. The theoretical results are given by Eq. (\ref{eq:popt1}).} 
\label{fig:popt}
\end{center}
\end{figure}

\begin{figure}[htbp]
\centering
\includegraphics[width=10cm,clip]{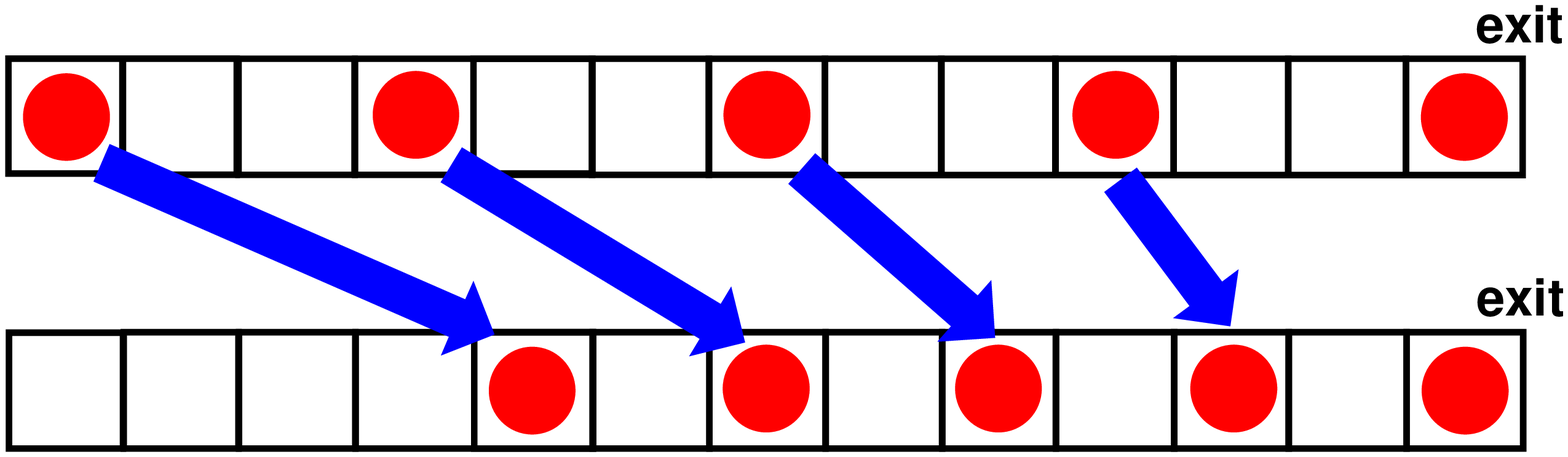}
\caption{The upper panel describes the configuration near the exit at $t=nT+\tau$ in our assumption, whereas the lower one describes the ideal configuration at $t=(n+1)T$. The particles further from the exit have to move more sites as in this figure.}
\label{fig:comment2-27}
\end{figure}

Figure \ref{fig:popt} plots the simulated and theoretical $p_{\rm opt}$ as a function of $T$ with red bars. The red bars show the range of $p$ which achieves more than 95\% of the improvement with $p_{\rm opt}$. Note that the detailed calculation scheme is described in \ref{sec:appendix}. 
In general, the analytical line shows the same trend as the simulation results, although the simulation results are less (higher) than the analytical line when $T<25$ $(T>25)$. These phenomena are explained as follows.

First, we consider the reason why simulation results are less than the analytical line when $T<25$. Basically, in our approximation, we assume that $p$ is a deterministic value rather than a probability. However, because $p$ is a probability in practice, the number of hops performed by one particle in one closing period have some variance. This variance influences the flow as follows; if the actual number of sites which each particle hops in one closing period becomes larger than the average number ($=p_{\rm opt}(T-\tau)$), the SlS effect occurs near the exit (see Fig. \ref{fig:exit}) and deteriorates the flow. Therefore, particles have to hop with a smaller $p$ to decrease the possibility of the occurrence of the SlS effect. As a result, the simulated values of $p_{\rm opt}$ becomes less than the analytical line.



Meanwhile, when $T>25$, the configuration of particles far from the exit, which is neglected in our approximation, also influences on the flow. Note that this will be discussed further in Subsec. \ref{sec:e}. We calculate $p_{\rm opt}$ in our approximation considering only two particles near the exit. However, other particles behind the second particle have to hop with a larger $p$ in order to achieve an appropriate gap between particles, otherwise their gap exceeds more than one site and deteriorates the flow (see Fig. \ref{fig:comment2-27}). 
This is because we assume that particles are separated by a two-site gap at $t=nT+\tau$. Eventually, the optimal $p$ for other particles behind the second particle becomes larger, pushing up the simulation results of $p_{\rm opt}$ over the theoretical line.





\subsection{Improvements of $Q$ for various $s$}
\label{sec:c}

In actual situations, the values of $s$ depend on particles' property, i.e., the law of inertia and the reaction delay. Therefore, we investigate the change of $a$ varying $s$ in this subsection.

\begin{figure}[h]
\begin{center}
\includegraphics[width=9cm,clip]{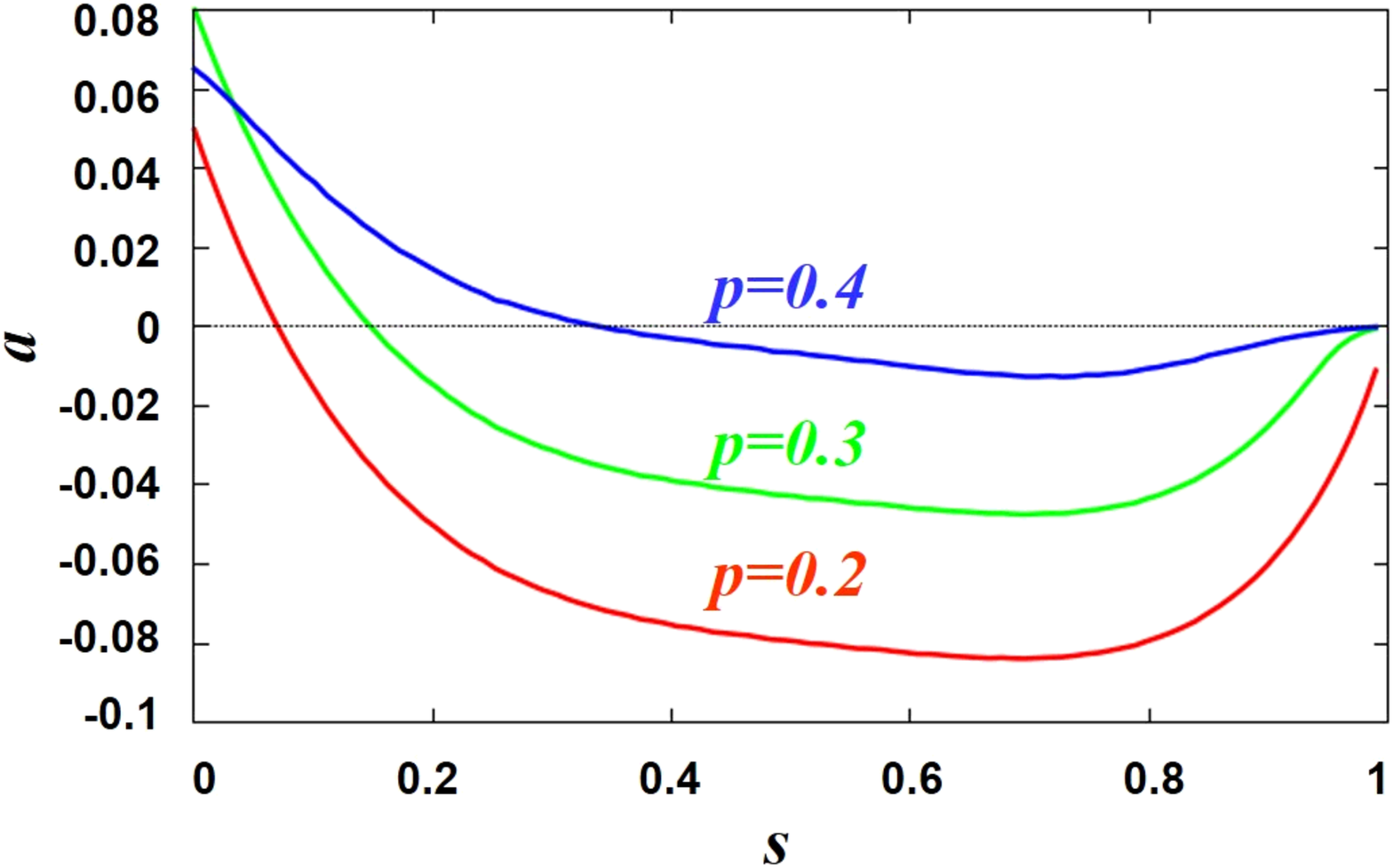}
\caption{Simulated values of $a$ as a function of $s$ for various $p$ $\in$ \{$0.2$ (red), $0.3$ (green), $0.4$ (blue)\}. The other parameters are set as $L=200$, $\alpha=1$, $\beta^*=0.6$, $T=20$, and $s=0$.}
\label{fig:sp}
\end{center}
\end{figure}

We plot the simulated values of $a$ obtained as a function of $s$ for $p$ $\in$ \{$0.2$ (red), $0.3$ (green), $0.4$ (blue)\} with $T=20$. The results are plotted in Fig. \ref{fig:sp}. Note that we write $Q$ and $a$, not $Q(p)$ and $a(p)$, in Subsec. \ref{sec:c}, \ref{sec:d}, and \ref{sec:e} because both $Q$ and $a$ depend also on other parameters. 

The control is effective when $s$ is relatively small, that is, when the SlS effect is relatively large, and becomes detrimental as $s$ increases. 
When $s$ is relatively large ($s>0.7$), $a$ recovers toward 0 (finally reaching $a=0$ at $s=1$). 
This phenomenon is explained as follows. In the region where $s>0.7$, the SlS effect is drastically weakened. Therefore, the flow is scarcely affected by the configuration of particles, and the control exerts little influence on the flow.

\subsection{Improvements of $Q$ for various $r$}
\label{sec:d}

In this subsection, we introduce a new important parameter, $r$. The parameter $r$ defines the proportion of particles which obey the control. Each particle entering the lattice obeys the control with probability $r$, and disobeys it with probability $1-r$. All particles in the lattice obey the control when $r=1$; conversely, no particles obey the control when $r=0$. 
The obedience or disobedience of each particle is determined at the left boundary and fixed while the particle resides on the lattice.

\begin{figure}[htbp]
\begin{tabular}{cc}
\begin{minipage}[c]{0.5\linewidth}
\flushleft
\includegraphics[keepaspectratio, height=5cm]{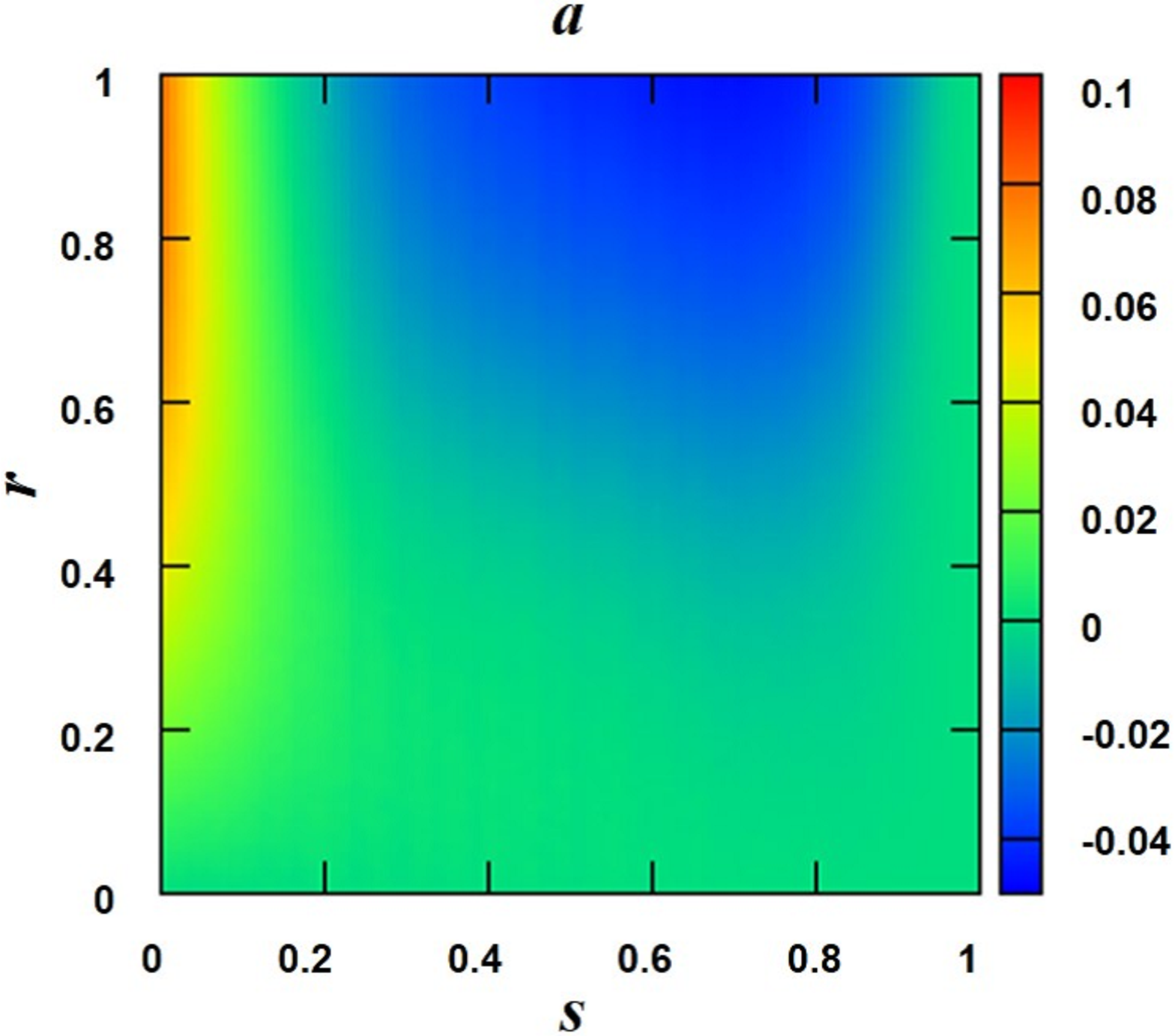}
\subcaption{}\label{fig:sr}
\end{minipage} &
\begin{minipage}[c]{0.5\linewidth}
\flushleft
\includegraphics[keepaspectratio, height=5cm]{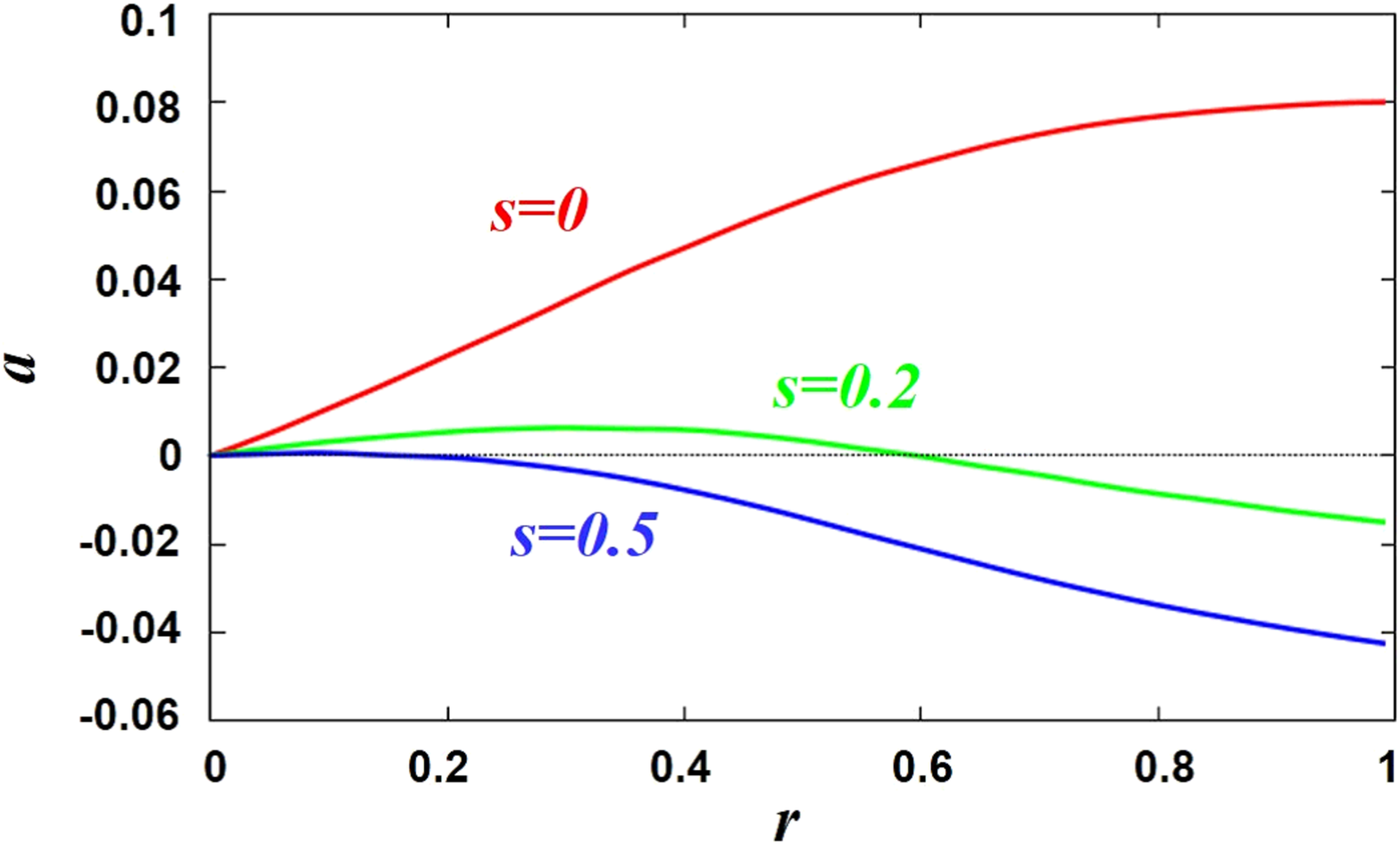}
\subcaption{}\label{fig:r}
\end{minipage}
\end{tabular}
\caption{(a) Simulated values of $a$ as a function of $s$ and $r$. The other parameters are set as $L=200$, $\alpha=1$, $\beta^*=0.6$, $T=20$, and $p=0.3$. (b) Simulated values of $a$ as a function of $r$ for $s$ $\in$ \{$0$, $0.2$, $0.5$\}. The other parameters are set as $L=200$, $\alpha=1$, $\beta^*=0.6$, $T=20$, and $p=0.3$.}
\end{figure}

Here, we investigate the coupled effects of $s$ and $r$ on $a$. Figure \ref{fig:sr} presents the simulated values of $a$ as a function of $s$ and $r$ with $T=20$ and $p=0.3$.
Between $s=0$ and $s\approx0.2$, we find that $a \geq 0$ for almost all the values of $r$. On the other hand, between $s\approx0.2$ and $s\approx0.9$, $a$ becomes negative when $r$ becomes large. Note that when $s\approx1$ (i.e., the SlS effect is almost 0), $a$ also diminishes to nearly 0, because the SlS effect becomes very weak and the control exerts little influence on the flow (see Subsec. \ref{sec:c}). 

To clarify the $r$-dependence of $a$, we plot $a$ versus $r$ for various values of $s$ $\in$ \{$0$, $0.2$, $0.5$\} in Fig. \ref{fig:r}.
In the case $s=0$ (indicating the strongest SlS effect), $a$ monotonically increases with $r$. As a result, the flow is maximized when all particles obey the control, i.e., $r=1$. Meanwhile, in the case $s=0.2$, the maximum $a$ is achieved at $r=0.32$. This result indicates that we should control an appropriate portion of particles as $s$ increases from 0. Finally, in the case $s=0.5$, $a$ monotonically decreases with $r$. Therefore, when the SlS effect is relatively small, the control can conversely deteriorate the flow.


\subsection{Optimal length of control}
\label{sec:e}
\begin{figure}[htbp]
\begin{center}
\includegraphics[width=9cm,clip]{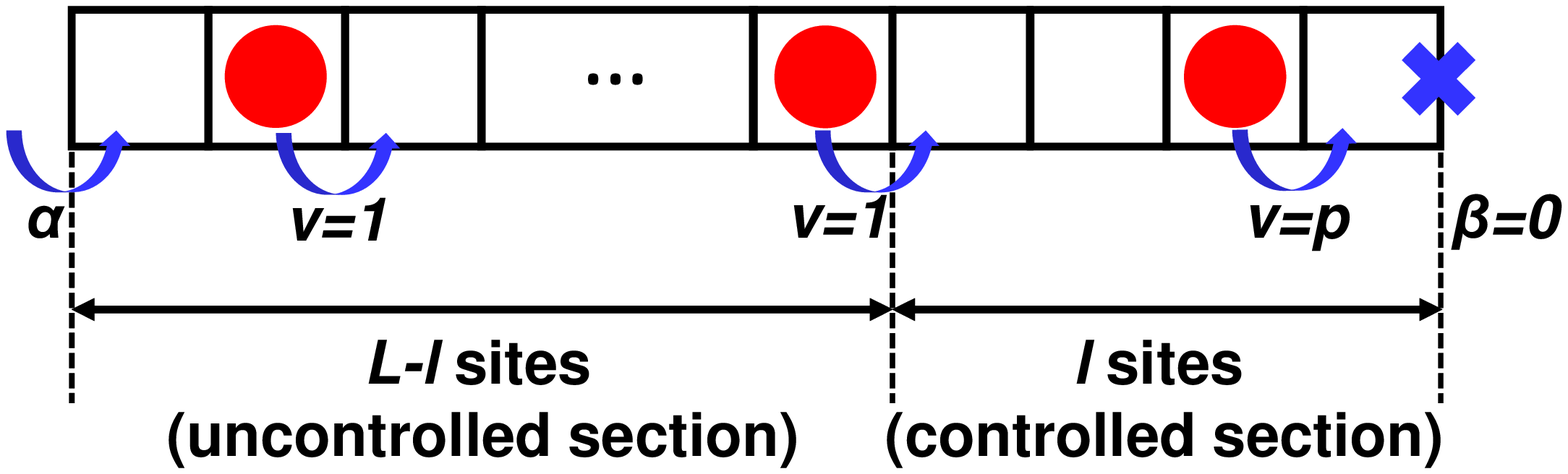}
\caption{Schematic definition of the controlled and the uncontrolled section during the closing period with $r=1$.}
\label{fig:l}
\end{center}
\end{figure}

Finally, we introduce another important parameter $l$ in this subsection. The parameter $l$ defines the length of the section from the exit, where the control works. The control is valid throughout the lattice when $l=L$, and completely invalid when $l=0$. Figure \ref{fig:l} describes the schematic of $l$.

In this subsection, we investigate the effect of $l$ on $a$. Note that we consider the most fundamental case, studied in Subsec. \ref{sec:a} and \ref{sec:b}, where $s=0$ and $r=1$.

\begin{figure}[htbp]
\begin{tabular}{cc}
\begin{minipage}[t]{0.5\linewidth}
\centering
\includegraphics[keepaspectratio, height=5cm]{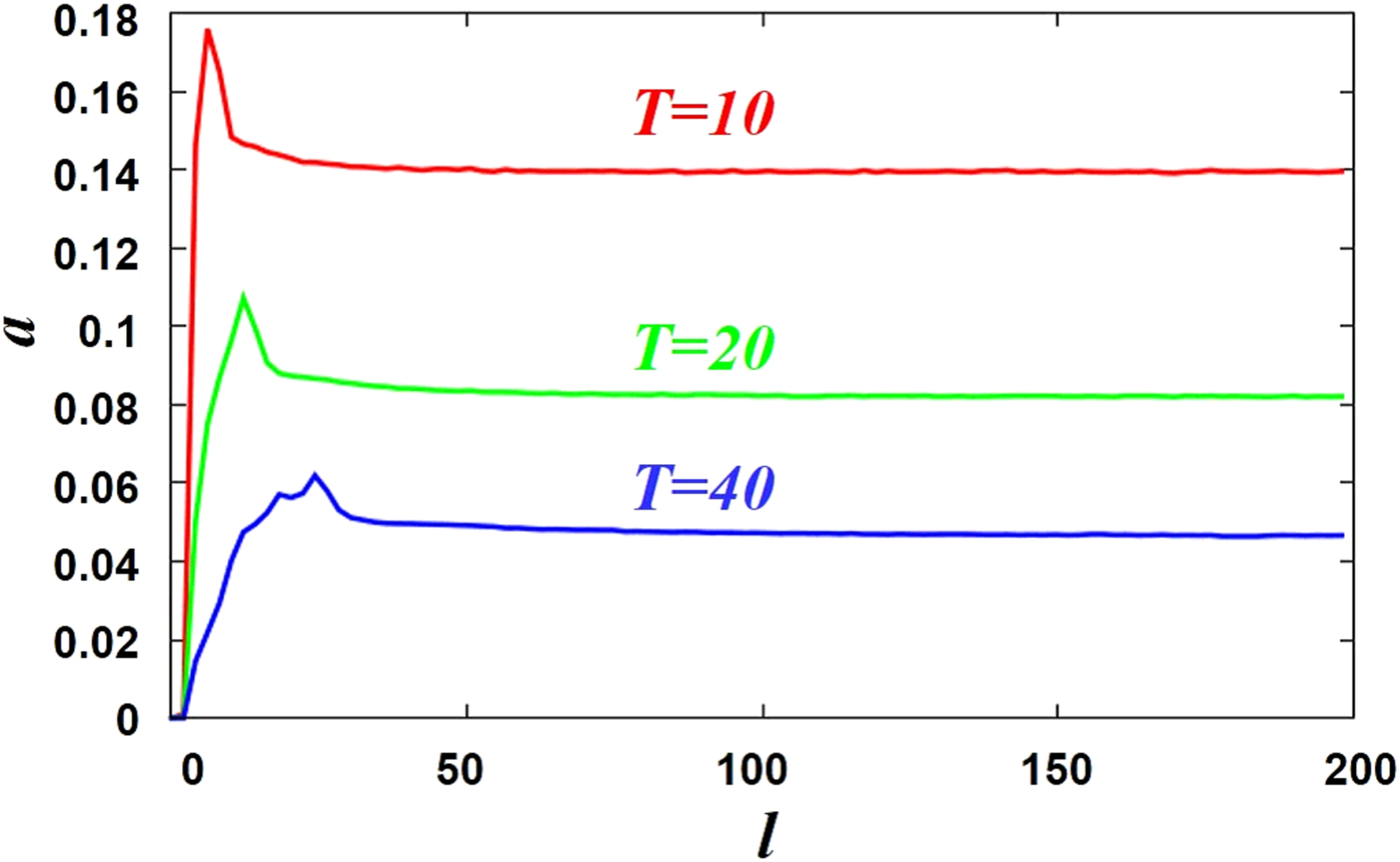}
\subcaption{}\label{fig:length}
\end{minipage} &
\begin{minipage}[t]{0.5\linewidth}
\centering
\includegraphics[keepaspectratio, height=5cm]{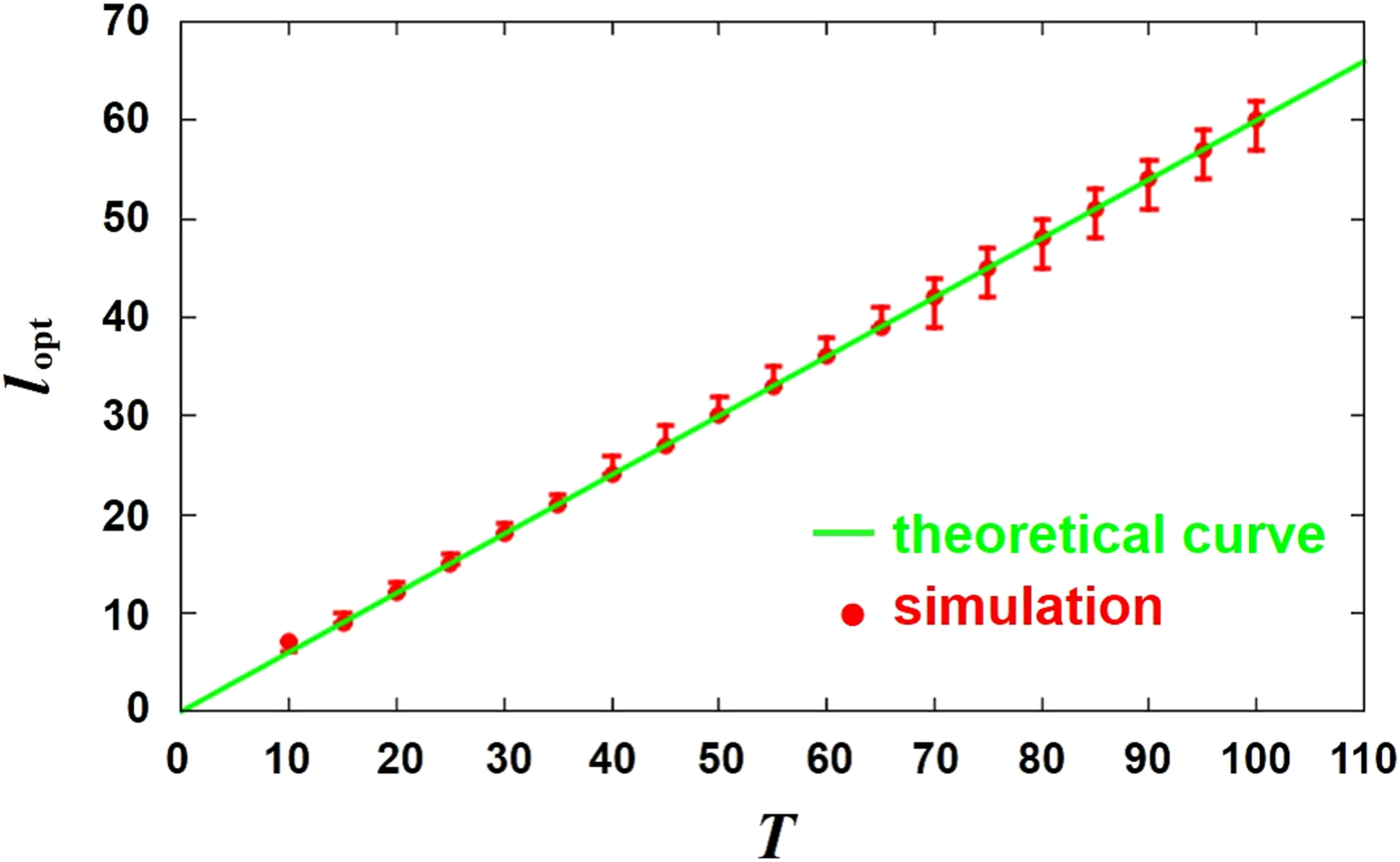}
\subcaption{}\label{fig:lopt}
\end{minipage}
\end{tabular}
\caption{(a) Simulated values of $a$ as a function of $l$ for $(T, p)\in$ \{(10, 0.47) (red), (20, 0.32) (green), (40, 0.21) (blue)\}. The other parameters are set as $L=200$, $\alpha=1$, $\beta^*=0.6$, $s=0$, and $r=1$. (b) Simulated (red dots) and theoretical (green line) values of $l_{\rm opt}$ as a function of $T$. The red bars show the range of $l$ which achieves more than 95\% of the improvement with $l_{\rm opt}$. The other parameters are set as $L=200$, $\alpha=1$, $\beta^*=0.6$, $s=0$, and $r=1$. The theoretical results are obtained by Eq. (\ref{eq:lopt1}).}
\end{figure}

Figure \ref{fig:length} plots the simulated values of $a$ as a function of $l$ for various $T \in \{10 ,20, 40\}$. Here, referring to Fig. \ref{fig:popt}, we set $p$ to its $p_{\rm opt}$ at each value of $T$, obtained by the simulations (see Fig. 10). Specifically, for $T=$ 10, 20, and 40 in Fig. \ref{fig:length}, we set $p$ = 0.47, 0.32, and 0.21, respectively.
In Fig. \ref{fig:length}, we observe that $a$ is maximized at some optimal value of $l$ ($l_{\rm opt}$) depending on $T$. Specifically, we find that $l_{\rm opt}(T=10)=7$, $l_{\rm opt}(T=20)=12$, and $l_{\rm opt}(T=40)=24$. Therefore, we conjecture that

\begin{equation}
l_{\rm opt}\approx1\times\beta^* T,
\label{eq:lopt1}
\end{equation}
where 1 denotes the hopping probability during an open period.

Figure \ref{fig:lopt} compares the simulated $l_{\rm opt}$ with red bars and the predictions of Eq. (\ref{eq:lopt1}) as a function of $T$. The red bars show the range of $l$ which achieves more than 95\% of the improvement with $l_{\rm opt}$. Note that the detailed calculation scheme is described in \ref{sec:appendix}.

The simulation results are in excellent agreement with Eq. (\ref{eq:lopt1}). 
To explain this phenomenon, we focus on the fact that particles located at most $1\times\beta^* T$ sites from the exit at $t=nT$, when $\beta$ is switched from 0 to 1, might leave the lattice during the next opening period. For example, the particle occupying the ($L-1\times\beta^* T$)th site at $t=nT$ can exit the lattice at $t=nT+\tau$ unless it is blocked by the leading particle during that open period. The number of particles leaving the lattice at $t=nT$ depends on the configuration of particles within $1\times\beta^* T$ sites from the exit during that period. Conversely, particles more than $1\times\beta^* T$ sites from the exit can never reach the exit within that period. Therefore, at the sites further than $1 \times \beta^* T$ sites from the exit, the control is meaningless and can be rather detrimental.


\section{Conclusion}
The present paper analyzes an effective control method, which improves the flow over a lattice with a bottleneck. The flow is enhanced by appropriately changing particles' hopping probability on the lattice, depending on the state of the bottleneck. Specifically, when $\beta=0$ ($\beta=1$), the hopping probability is set to $v=p \in (0,1)$ ($v=1$). Note that our model differs from the related works~\cite{1742-5468-2014-10-P10019, 1742-5468-2009-02-P02012, 1742-5468-2012-05-P05008, 1742-5468-2008-06-P06009,PhysRevE.87.062818}, in which the hopping probability is fixed and other properties, such as the input probability (rate), are altered.

Herein, we report a number of important results. 
In Subsec. \ref{sec:a} and \ref{sec:b}, we observed that the control improves the flow if $p$ is appropriately chosen. We also found that an optimal value of $p$ ($p_{\rm opt}$) which maximally improves the flow in the HD phase (without the control) depends on $T$. Specifically, the value of $p_{\rm opt}$ and the maximal value of $a(p)$ decreases if $\tau$ or $\beta^*$ is constant. Conversely, in the LD phase (without the control), the control impedes the flow. 
In Subsec. \ref{sec:c} and \ref{sec:d}, we investigated the influence of $s$ and $r$ on the flow. The flow was improved only when the SlS effect was relatively large. Especially, in the case where the SlS effect is maximum $(s=0)$, the flow improvement became the largest when all particles obeyed the control ($r=1$). As the SlS effect weakened (i.e., $s$ increased), the optimal $r$ diminished and finally reached 0 (indicating loss of the control). 
Finally, in Subsec. \ref{sec:e}, we found an optimal control length $l_{\rm opt}$ that maximizes the flow. Similar to $p_{\rm{opt}}$, $l_{\rm opt}$ is also $T$-dependent. 

According to these results, the effect of our control is maximized if we select an appropriate hopping probability and controlled section when the lattice is congested and the SlS effect is relatively large. 
We stress again that the essence of the results lies in that the flow improves even if deceleration by $p$ is stochastic, although the most ideal control is to make all particles keep an appropriate gap deterministically during deceleration.

In vehicular traffic contexts, it is reported that traffic flow through a bottleneck can be improved by appropriately decelerating vehicles. One of the proposed methods is variable speed limit (VSL) ~\cite{papageorgiou2008effects} control, where the limited speed of vehicles is changed according to the state of a bottleneck.
It is presumable to interpret our control as a sort of VSL control. Applying the results in this study for an actual traffic flow through a traffic light, we can show that our VSL control improves the flow by approximately 4\% in the following conditions: $T\approx20$ sec., $\tau\approx$12 sec., $v\approx$ 60 km/h (the maximal speed) or 12 km/h (the limited speed during a red-light period), and $l\approx$180 m. We note that the length of one site and the length of one time step are set as 7.5 m~\cite{nishi2009achievement} and 0.5 s, respectively.

We admit that more refined models are required to study a real traffic accurately. Such approaches, for example, introducing inhomogeneity of particles, remain as future works. However, we would like to mention that our simple model allows us to discuss the effect of VSL control as a first step.


\section*{Acknowledgements}
The authors greatly thank Takahiro Ezaki for helpful discussions and beneficial comments. This work was supported by JSPS KAKENHI Grant No. 25287026 and 15K17583.

\appendix

\section{
Dependence on $\tau$}
\label{sec:appendixt}
In this appendix, we vary the values of $\tau (=\beta^*T) \in \{10, 11, 12\}$ and investigate the improvements of $Q(p)$ to the model, fixing $L=200$, $\alpha=1$, and $T=20$. We note that $m$ ($m \in \mathbb{N}$) particles can leave the lattice with $s=0$ for $\tau \in \{3m-2, 3m-1, 3m\}$ in one open period, i.e., $Q(1)=0.2$ for $\tau \in \{10, 11, 12\}$ with $s=0$.

\begin{figure}[htbp]
\begin{tabular}{cc}
\begin{minipage}[c]{0.5\linewidth}
\centering
\includegraphics[keepaspectratio, height=5cm]{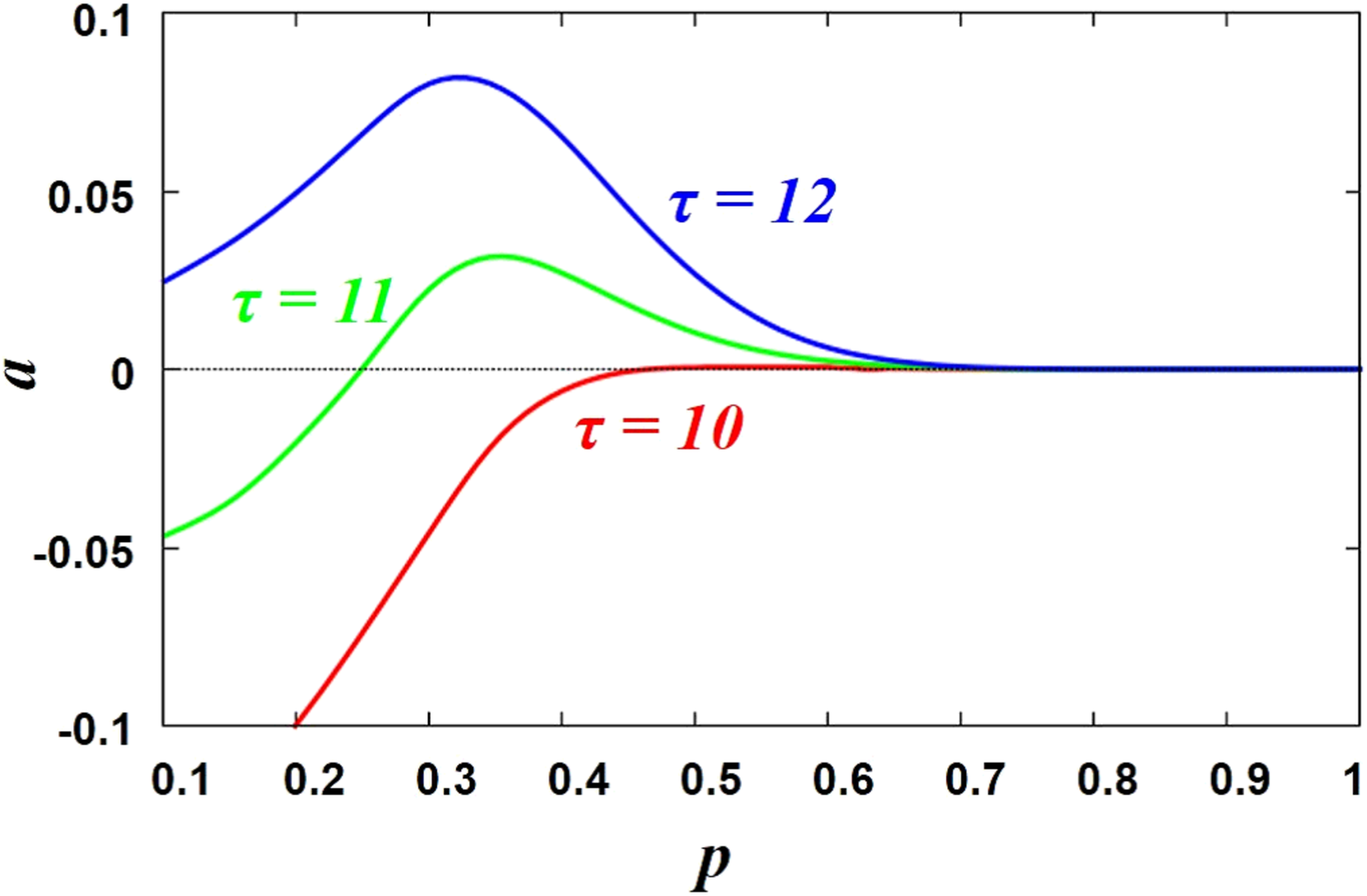}
\subcaption{}\label{fig:2-17-1}
\end{minipage} &
\begin{minipage}[c]{0.5\linewidth}
\centering
\includegraphics[keepaspectratio, height=5cm]{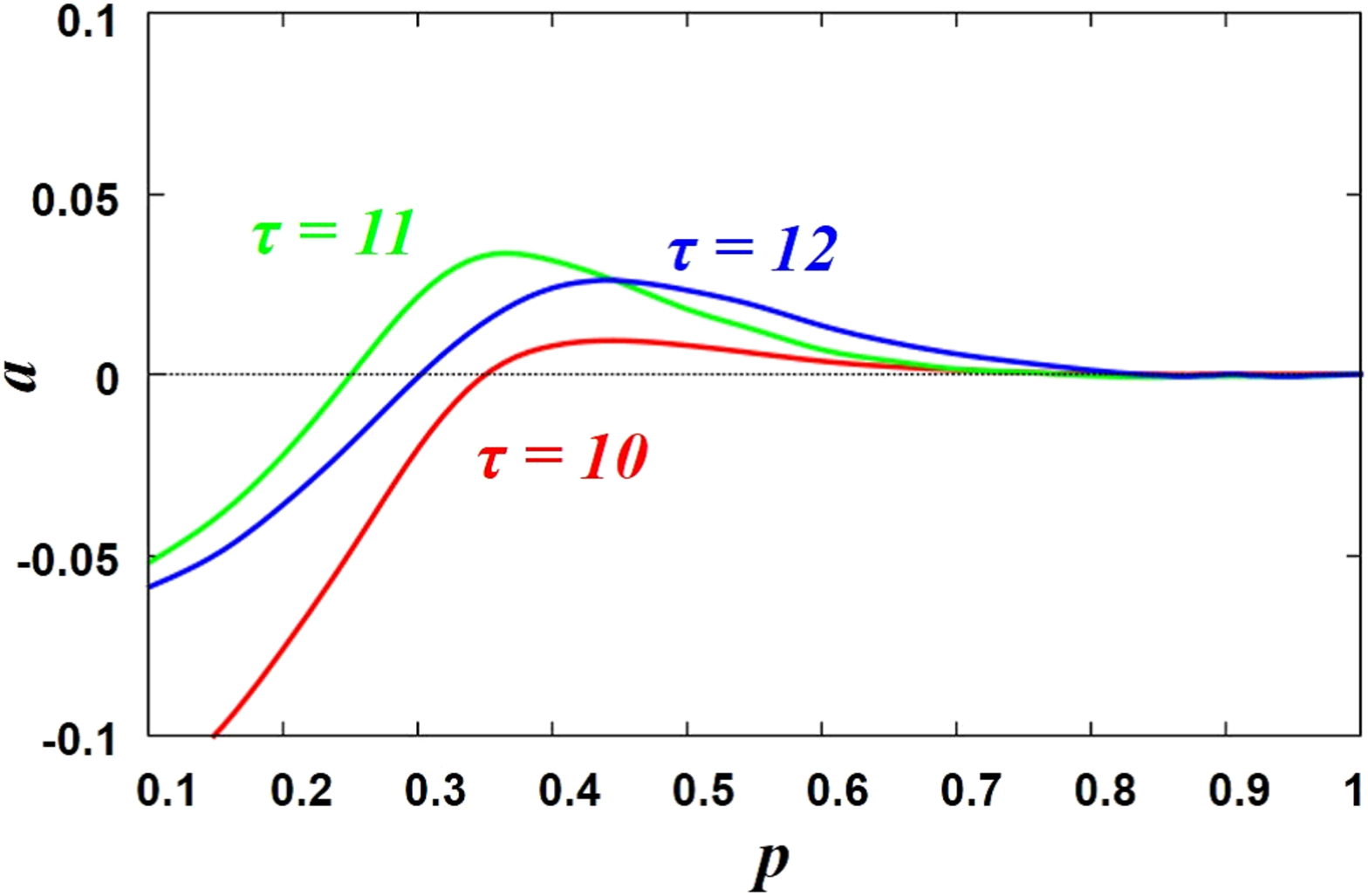}
\subcaption{}\label{fig:2-17-2}
\end{minipage}
\end{tabular}
\caption{Simulated values of improvement ratio $a$ as a function of $p$ for various $\tau \in \{10, 11, 12\}$ with (a) $s=0$ and (b) $s=0.15$. The other parameters are set as $L=200$, $\alpha=1$, and $T=20$.}
\label{fig:tau}
\end{figure}

The simulated values of improvement ratio $a(p)$ are plotted in Fig. \ref{fig:tau} for (a) $s=0$ and (b) $s=0.15$.

In Fig. \ref{fig:2-17-1}, the control improves the flow for $\tau \in \{11, 12\}$ if we appropriately select $p$, whereas the control never improves the flow for $\tau=10$. In the case of $\tau=3m-2(=10)$, there exists little room for flow improvement by our control. This is because $\tau=3m-2(=10)$ is the minimal time allowing $m(=3)$ particles to leave the lattice with $s=0$.

On the other hand, in Fig. \ref{fig:2-17-2}, although the improvements are clearly lessened for $\tau=11, 12$ compared to the left panel, the flow improves by 1 \% at most for $\tau=10$.

From the above discussion, we conclude that the effect of the control is (i) large but greatly depends on $\tau$ when $s=0$ and (ii) small but the dependence on the selection of $\tau$ gradually disappears as $s$ becomes large.

\section{
Dependence on SlS rules}
\label{sec:appendixs}

\begin{figure}[h]
\begin{tabular}{cc}
\begin{minipage}[t]{0.5\linewidth}
\centering
\includegraphics[keepaspectratio, height=5cm]{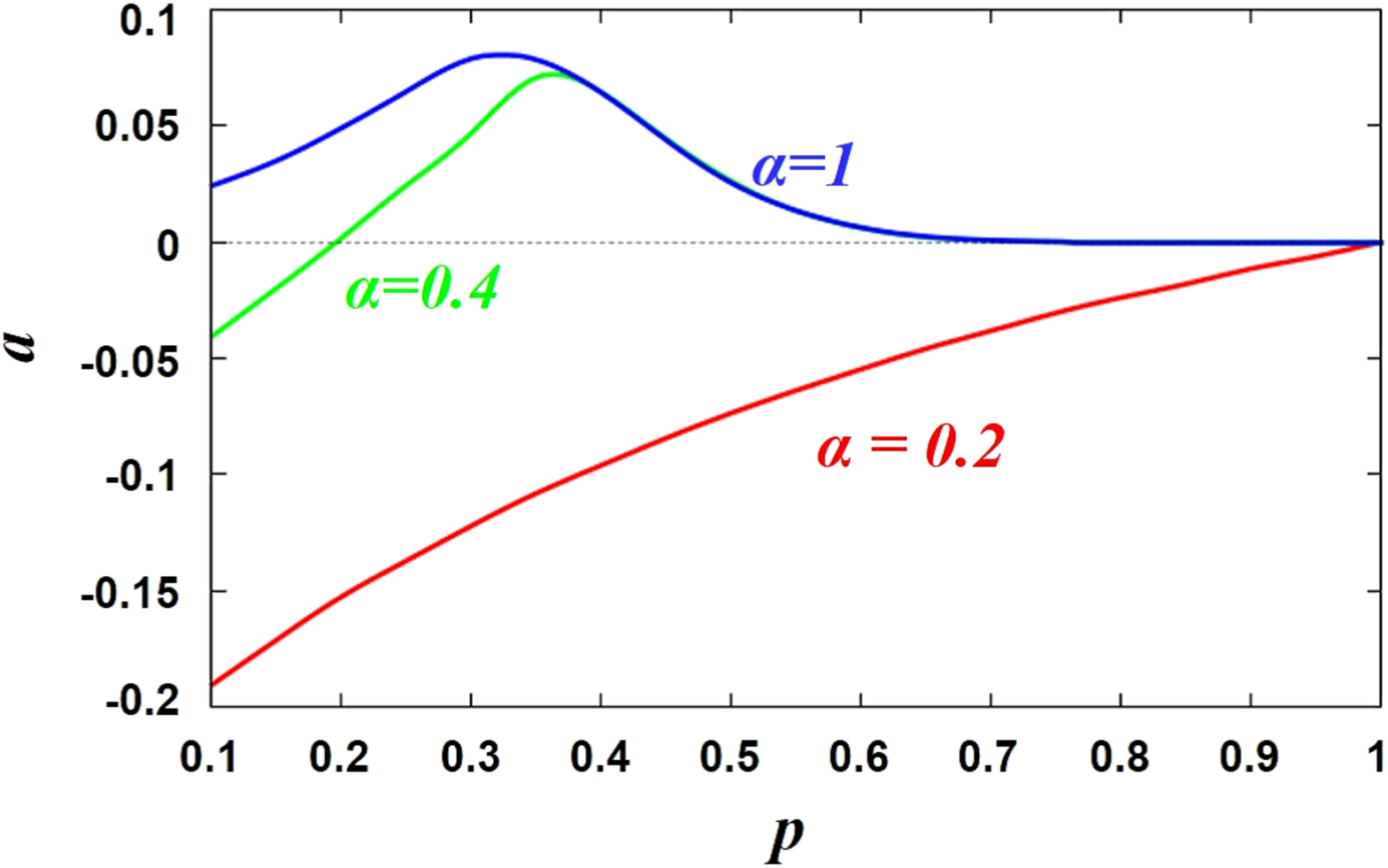}
\subcaption{}\label{fig:2-1-1}
\end{minipage} &
\begin{minipage}[t]{0.5\linewidth}
\centering
{\includegraphics[keepaspectratio, height=5cm]{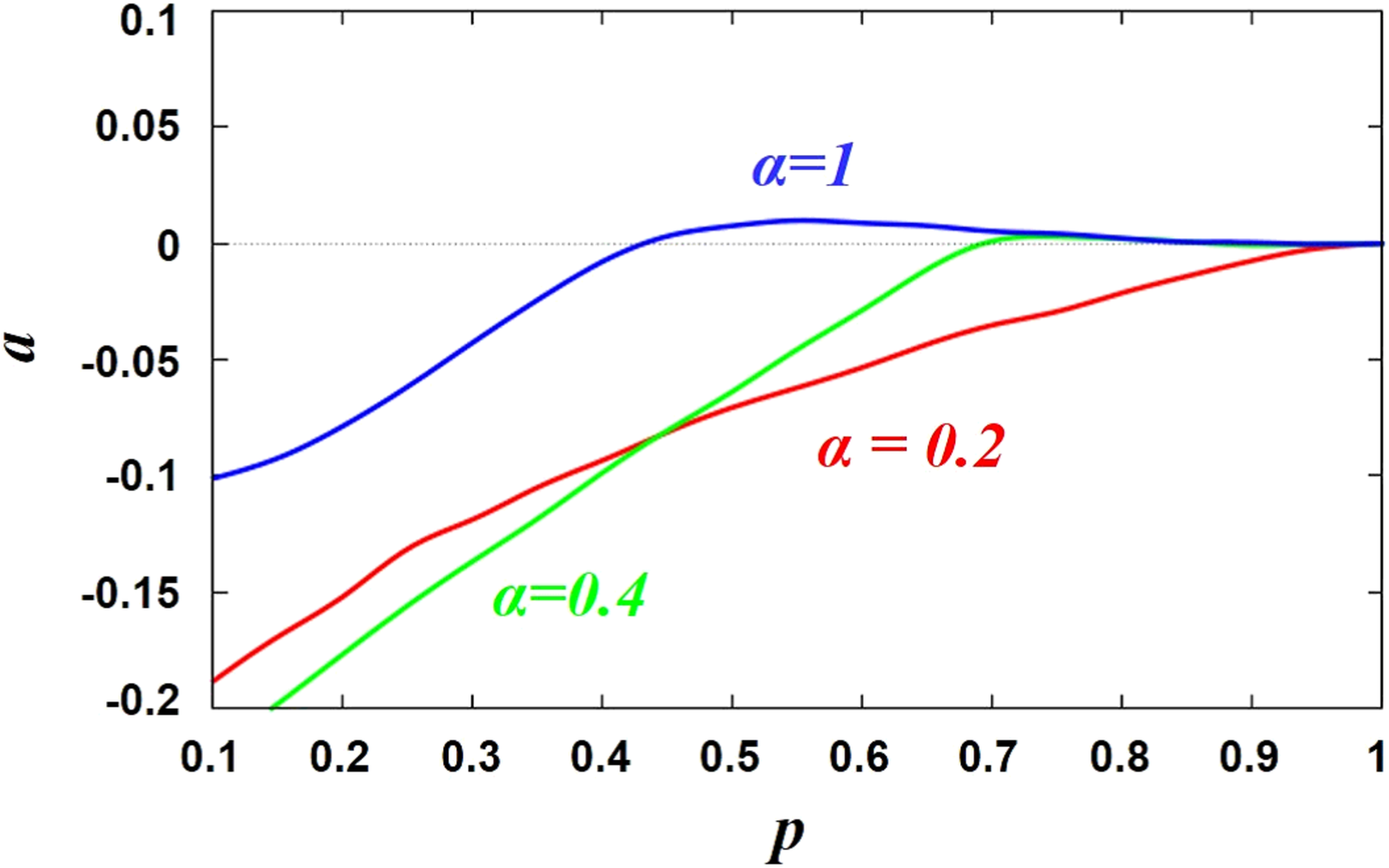}}
\subcaption{}\label{fig:2-1-2}
\end{minipage} 
\end{tabular}
\caption{Simulated values of $a(p)$ as a function of $p$ for various $\alpha$ $\in$ \{$0.2$(red), $0.4$(green), $1$(blue)\}. The other parameters are set as $L=200$, $\beta^*=0.6$, $T=20$, and (a) $(s_{\rm c}, s_{\rm o})=(0.5, 0)$ and (b) $(s_{\rm c}, s_{\rm o})=(0, 0.5)$.}
\label{fig:2-1}
\end{figure}

Throughout the paper, we assume a constant $s$. In this appendix, we vary $s$ dependent on $\beta$. Specifically, $s=s_{\rm c}$ during close periods and $s=s_{\rm o}$ during open periods, where $s_{\rm o} \not = s_{\rm_c}$. We describe two figures corresponding to Fig. \ref{fig:kaizen1}, setting (a) $(s_{\rm c}, s_{\rm o})=(0.5, 0)$ and (b) $(s_{\rm c}, s_{\rm o})=(0, 0.5)$. The simulation results are described in Fig. \ref{fig:2-1}.

Fig. \ref{fig:2-1-1} almost coincides with Fig. \ref{fig:kaizen1}. On the other hand, flow improvement is greatly decreased in Fig. \ref{fig:2-1-2} due to a larger $s$ during open periods. Therefore, we conclude that the SlS coefficient during open periods greatly affects the flow in our model.

Note that our SlS rule is basically based on Benjamin-Johnson-Hui (BJH) model~\cite{0305-4470-29-12-018}. There are other typical SlS rules, which are implemented in Takayasu and Takayasu ($\rm T^2$) model~\cite{doi:10.1142/S0218348X93000885} and Appert-Rolland and Santen (AS) model~\cite{PhysRevLett.86.2498}. In the former, a blocked particle cannot hop until its leading particle stay two-sites ahead of it, whereas in the latter the SlS effect remains until it finally hops. Both models have stronger SlS effects than BJH model. Therefore, we are convinced that our control can have a positive effect on the flow change if our SlS rule is modified based on either $\rm T^2$ or AS models.

\section{
Implementation of Woelki's method}
\label{sec:appendixw}
In Woelki's approach, input rate is decreased (increased) if the average density of the system exceeds (falls below) a threshold density $\rho_t$. Noting the difference from Woelki's approach, we focus on the control of a hopping probability (particle's velocity), while Woelki focused on the control of an input rate. Moreover, the SlS effect is not considered in his work.

\begin{figure}[h]
\begin{center}
\includegraphics[width=9cm,clip]{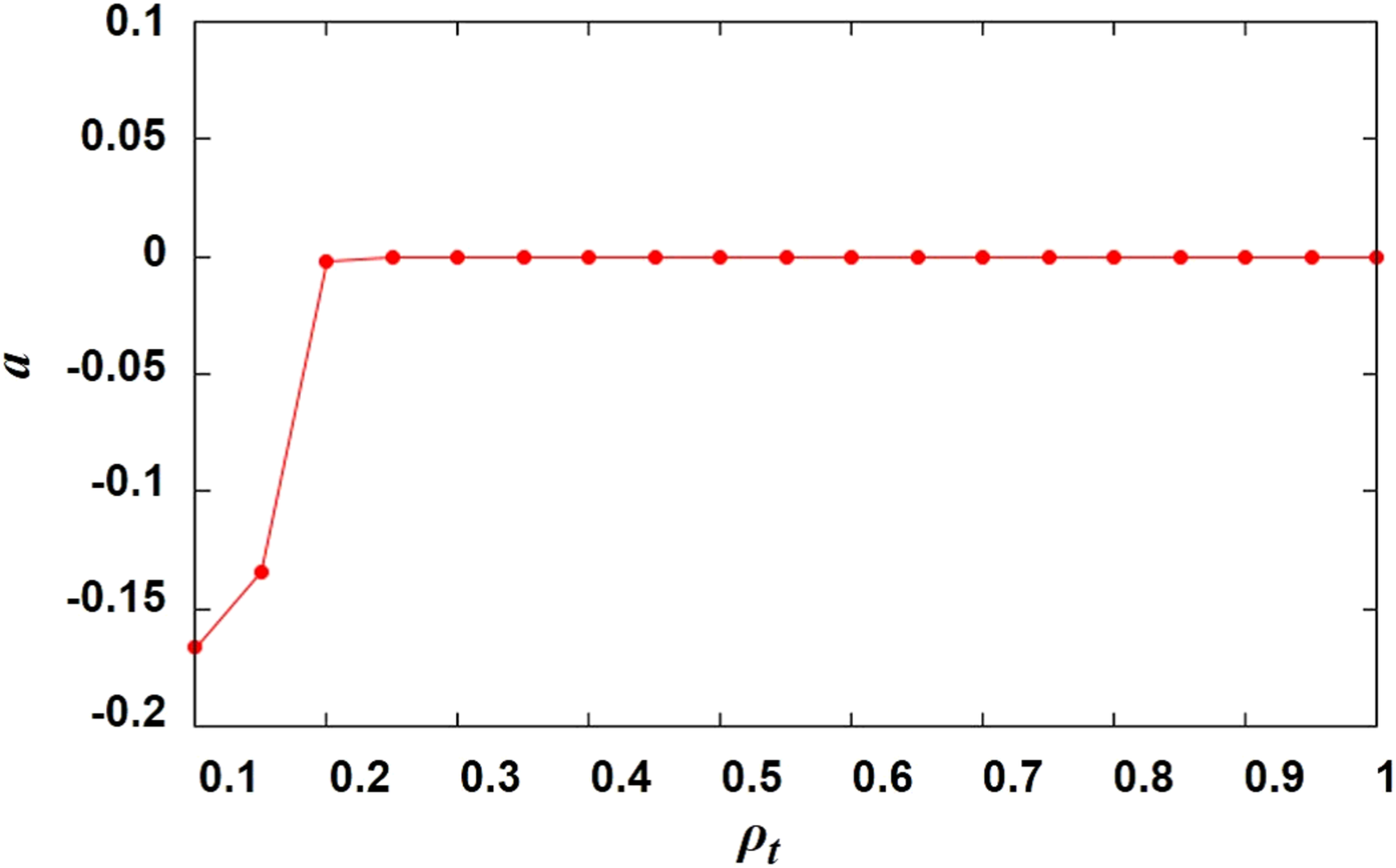}
\caption{Improvement ratio $a$ as a function of the threshold density $\rho_{\rm \rm t}$ with Woelki's control rule. The other parameters are set as $L=200$, $\alpha=1$, $\beta^*=0.6$, and $T=20$. Changing only the global density of the lattice is ineffective in the TASEP with the SlS rule.}
\label{fig:woelki}
\end{center}
\end{figure}

As a result of simulations of our model applied with Woelki's rule, the flow is not improved if we control input probability depending on the global density as with Woelki's approach. Figure \ref{fig:woelki} shows the results of simulations by Woelki's rule, where the horizontal axis represents threshold density $\rho_{\rm t}$ and the vertical axis represents improvement ratio of the flow $a$. We observe that there is no flow improvement ($a>0$) for any $\rho_{\rm t}$.

In conclusion, when we introduce the SlS rule, the configuration of particles, which is controlled by our method, is important for flow improvement rather than the global density of the lattice, which is controlled by Woelki's control.

\section{Calculation of $p_{opt}$ ($l_{opt}$)}
\label{sec:appendix}
Here, we explain how we pick up $p_{opt}$ ($l_{opt}$) and depict red bars in Fig. \ref{fig:popt} and Fig. \ref{fig:lopt}.
First, we calculate the improvement ratio $a$ in increments of $\Delta p=0.01$ ($\Delta l=1$) and determine the value of $p$ ($l$) which achieve the maximum $a$ as $p_{\rm opt}$ ($l_{\rm opt}$). Then, we extract the values of $p$ ($l$) which achieves more than 95\% of the improvement ratio $a$ with $p_{\rm opt}$ ($l_{\rm opt}$) and depict red bars.

\section*{References}
\providecommand{\newblock}{}

\end{document}